\documentclass{article}

    \PassOptionsToPackage{numbers, compress}{natbib}
\usepackage[main, final]{neurips_2026}

\usepackage[utf8]{inputenc} % allow utf-8 input
\usepackage[T1]{fontenc}    % use 8-bit T1 fonts
\usepackage{hyperref}       % hyperlinks
\usepackage{url}            % simple URL typesetting
\usepackage{xurl}
\usepackage{booktabs}       % professional-quality tables
\usepackage{amsfonts}       % blackboard math symbols
\usepackage{nicefrac}       % compact symbols for 1/2, etc.
\usepackage{microtype}      % microtypography
\usepackage{xcolor}         % colors

\definecolor{linkblue}{RGB}{0,70,140}
\definecolor{citegreen}{RGB}{0,110,60}
\hypersetup{
  colorlinks = true,
  linkcolor  = black,   % Fig./Sec./Eq./Appendix refs
  citecolor  = black,  % [12] citations
  urlcolor   = black,    % URLs (code link in App. A)
    pdftitle={Growing a Neural Network in Breadth, Depth, and Time},
    pdfauthor={Eivinas Butkus, Kedar Garzón Gupta, Nikolaus Kriegeskorte}
}

\usepackage{amsmath}
\usepackage{graphicx}
\usepackage{comment}
\usepackage{subcaption}
\usepackage{tikz}

\title{Growing a Neural Network\\ in Breadth, Depth, and Time}

\author{
\begin{tabular}{c@{\hspace{0.5cm}}c@{\hspace{0.5cm}}c}
Eivinas Butkus$^{1,2}$ & Kedar Garzón Gupta$^{1}$ & Nikolaus Kriegeskorte$^{1,2}$ \\
\texttt{eb3407@columbia.edu} &
\texttt{kg3162@columbia.edu} &
\texttt{nk2765@columbia.edu}
\end{tabular} \\[1.5em]
$^1$Columbia University \\
$^2$NSF AI Institute for Artificial and Natural Intelligence
}

\newcommand{\lambreadth}[0]{\lambda_{\text{breadth}}}
\newcommand{\lamdepth}[0]{\lambda_{\text{depth}}}
\newcommand{\lamtime}[0]{\lambda_{\text{time}}}

\usepackage{enumitem}
\usepackage{multirow}

\usepackage{newfloat}
\DeclareFloatingEnvironment[
  fileext=losf,
  placement={htbp},
  name=Figure
]{suppfigure}

\usepackage{float}

\begin{document}

\maketitle

\begin{abstract}
    Brains and AI systems alike must achieve their goals with limited space, time, energy, and data.
    Here we focus on space and time, defining differentiable costs for breadth (channels), depth (layers), and time (recurrent steps) in a recurrent convolutional network.
    We optimize these costs jointly with task error via backpropagation.
    Pricing each resource differently, we grow a diverse set of computational graphs through optimization.
    We find that breadth, depth, and time can be traded off against each other at a given level of accuracy.
    Networks grow with task complexity and, when allowed to adaptively allocate time per input, take more recurrent steps if inputs are occluded.
    The time the model spends on each image also correlates with human reaction times, despite no training on human data.
    Our work takes a step toward a normative account of how spatial and temporal resource costs jointly shape neural architectures in brains and machines.
\end{abstract}

% main text

\section{Introduction}\label{section:introduction}

Intelligence---both biological and artificial---can be broadly characterized as the capacity to achieve goals with limited space, time, energy, and data \cite{simon1955behavioral, gershman_computational_2015, sterling_principles_2015, lieder_resource-rational_2020}.
Two of these resources, \textit{space} and \textit{time}, are the focus of this work.
In brains, each additional neuron adds volume, wiring, and maintenance costs, so, all else being equal, a smaller brain is preferable \cite{chklovskii_wiring_2002, laughlin2003communication, sterling_principles_2015}.
A faster brain confers advantages too.
An animal that is slow to detect a predator may not survive.
To understand a brain fully, we must therefore consider not only the problems it solves but also the spatial and temporal limitations under which it evolved.

Neural network models have become a central tool for studying how brains compute \cite{yamins2014performance,doerig2023neuroconnectionist,hasson2020direct}.
Most, however, optimize task performance alone or engage a single resource, e.g., space, through spatial constraints that explain functional organization in cortex \cite{blauch2022connectivity,margalit2024unifying}, or time, through recurrent processing that captures the dynamics of visual inference \cite{kietzmann2019recurrence,spoerer2020recurrent}.
Yet resources interact.
Could a smaller visual system match a larger one by running longer?
Under time pressure, is a wider or a deeper hierarchy better?
In the brain, each processing step takes a few milliseconds, leaving room for only about a hundred serial steps in a second-long cognitive process, favoring massive parallelism \cite{thorpe1996speed,feldman1985connectionist,rumelhart1986general,suzuki2023deep}, yet costly neurons and connections favor a compact circuit reused over multiple recurrent steps \cite{kubilius2019brain,nayebi2022recurrent}.
Biological neural systems must balance these opposing pressures.

AI systems face similar pressures.
A vision model on a phone or in a self-driving car must fit in limited memory, respond in real time, and conserve power.
Work on model compression \cite{han_deep_2016}, distillation \cite{hinton2015distilling}, architecture search \cite{zoph2017neural, liu_progressive_2018, cai2019proxylessnas}, and adaptive computation \cite{raposo2024mixture, snell2024scaling} mostly targets a single resource or trades resources by hand \cite{tan2019efficientnet}, and sweeping fixed architectures over several dimensions scales poorly.
Mapping the joint landscape of space, time, and performance directly would reveal which combinations are efficient for a given budget, including on neuromorphic hardware, where costs on activity and connectivity may track the physical cost of computation more closely \cite{kudithipudi2025neuromorphic}.

\begin{figure}[t]
    \centering
    \begin{tikzpicture}
        \node[anchor=south west, inner sep=0] (img) at (0,0)
            {\includegraphics[trim={0.2cm 14cm 0.2cm 0.2cm}, clip, 
             width=0.85\linewidth]{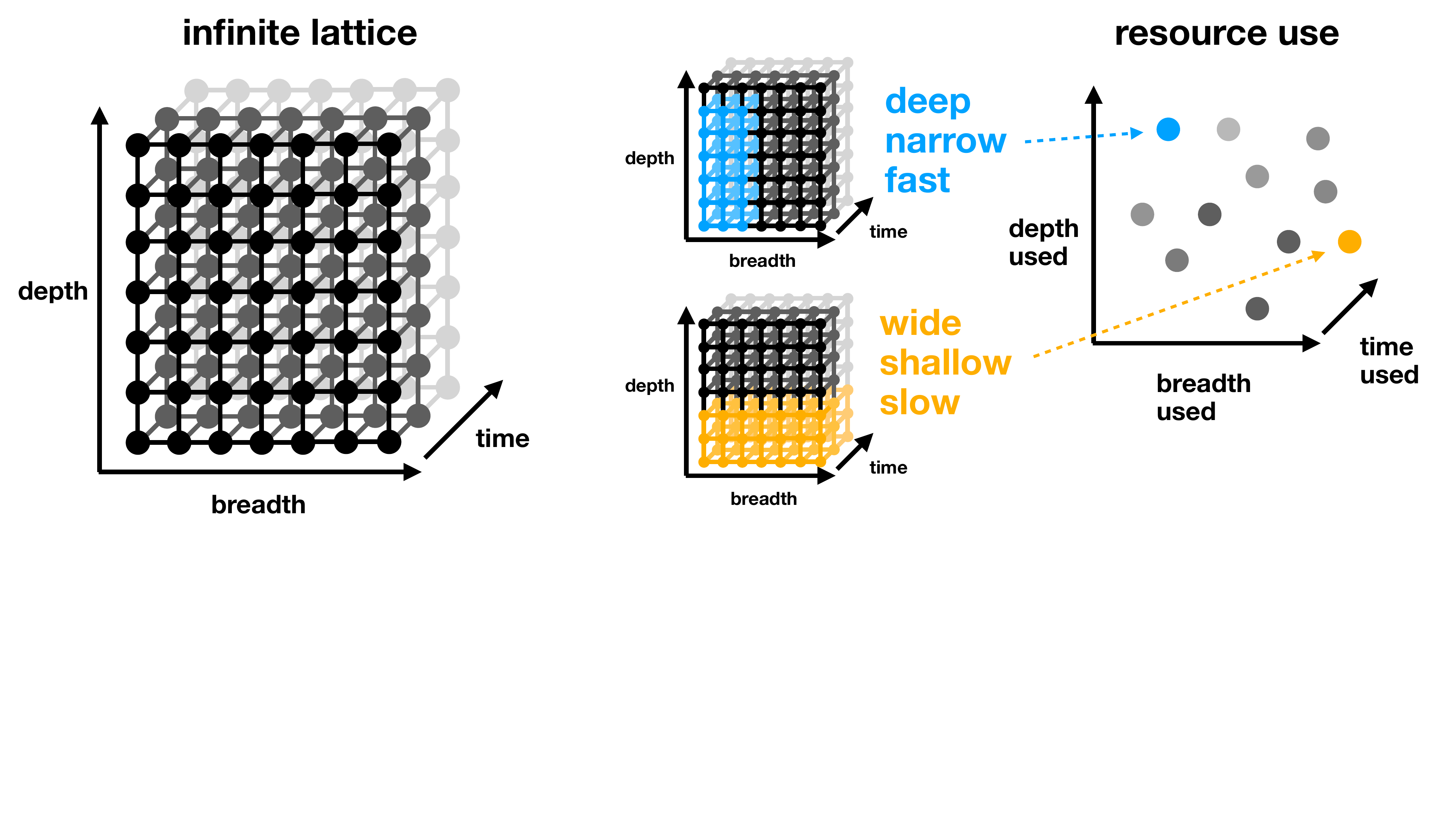}};
        \begin{scope}[
            x={(img.south east)}, 
            y={(img.north west)}
        ]
            \node[font=\bfseries] at (0.02, 0.97) {a};
            \node[font=\bfseries] at (0.4, 0.97) {b};
        \end{scope}
    \end{tikzpicture}
    \caption{
    \textbf{Growing a neural network in an infinite lattice.}
    ~\textbf{a}~
    The space of possible computational graphs can be pictured as an infinite lattice whose axes are resources (here breadth, depth, and time).
    ~\textbf{b}~
    Each model instance occupies a finite subset of the lattice with its own resource profile, e.g., deep, narrow, and fast (blue) or wide, shallow, and slow (orange).
    Optimizing differentiable resource costs lets each model instance find its own position in the resource-use space (right).   
    }
    \label{fig:lattice-solutions}
\end{figure}

In this work, we consider three resources---breadth, depth, and time---in recurrent convolutional neural networks.
Breadth and depth are spatial resources, corresponding to the number of channels and layers used, and time corresponds to the number of recurrent processing steps.
Prior work has optimized some of these individually.
For instance, pruning methods remove weights or channels post-hoc \cite{han_learning_2015, li_pruning_2017, frankle_lottery_2019}, network slimming learns channel importance via differentiable sparsity penalties \cite{liu_learning_2017}, and adaptive computation time optimizes time allocation through backpropagation \cite{graves_adaptive_2016}.
Here we instead define differentiable cost terms for breadth, depth, and time, and optimize them \textit{jointly} with task performance via backpropagation.
Rather than fixing the architecture in advance, this lets a network \textit{grow} in all three dimensions, finding its own trade-off between resource use and task performance (Fig.~\ref{fig:lattice-solutions}).
Optimization starts from an overcomplete graph, and the resource costs prune away what the task does not need, loosely mimicking brain development, where an initial growth to an overabundance of synapses is followed by selective pruning \cite{rakic1986concurrent,huttenlocher1997regional}.

We train over a thousand models under diverse resource pressures. Our main contributions are:

\begin{enumerate}[leftmargin=*,
labelindent=1em,
itemsep=2pt, topsep=2pt]
    \item A first instantiation of the multi-resource cost (MRC) approach, in which differentiable costs for breadth, depth, and time are optimized jointly with task error via backpropagation, and which can naturally be extended to other resources such as energy and data.
    \item To our knowledge, the first joint exploration of trade-offs between breadth, depth, and time, showing that all three resources can compensate for one another.
    \item The finding that adaptive time allocation emerges organically and outperforms fixed allocation.
    Models take more recurrent steps on occluded inputs, and their per-image time partly tracks human reaction times, even after controlling for image difficulty.
    \item The finding that networks use more resources with increasing task complexity.
\end{enumerate}

\section{Related work}\label{section:related-work}

% resource rationality and cognitive science
The idea that intelligent systems operate under resource limitations has a long history.
Simon's bounded rationality \cite{simon1955behavioral} proposed that decision-makers optimize within cognitive limits, an idea formalized more recently as computational rationality \cite{gershman_computational_2015} and resource rationality \cite{lieder_resource-rational_2020}.
Symbolic and probabilistic computational models have explored how resource-rational agents perceive and make decisions \cite{vul2014one, ho_people_2022, belledonne2026adaptive}.
These works study resource costs at the cognitive level, whereas we impose them at the level of neural architecture (on the wiring and dynamics of the network).

% cog comp neuro
In computational neuroscience, spatial resource constraints have been studied through the lens of wiring economy---the principle that neural circuits minimize wiring costs \cite{chklovskii_wiring_2002, chen_wiring_2006, achterberg_spatially_2023}.
Spatial constraints in neural network models have been used to explain early visual representations \cite{lindsey_unified_2019} and the topographic organization of visual cortex \cite{blauch2022connectivity, margalit2024unifying}.
On the temporal side, recurrent neural networks help explain the dynamics of human object recognition \cite{spoerer_recurrent_2017, kietzmann2019recurrence, spoerer2020recurrent}.
Recurrence is especially important for occluded or challenging images \cite{tang2018recurrent, kar2019evidence}, and the time recurrent models take to reach a confident decision can predict human reaction times \cite{spoerer2020recurrent, goetschalckx2023computing}.
Closest to our question, shallow recurrent networks can rival deeper feedforward ones on object recognition while better matching primate neural and behavioral data \cite{kubilius2019brain}, and recurrence in ventral-stream models appears to trade spatial for temporal complexity \cite{nayebi2022recurrent}.
These studies compared hand-designed architectures; our framework lets the trade-off between spatial and temporal resources emerge through optimization.

% deep learning
In deep learning, prior work has learned how much of a resource to use in two senses.
Some methods learn a static architecture size, such as the number of channels via sparsity penalties \cite{liu_learning_2017, alvarez2016learning, louizos2018learning} or the number of layers \cite{nazaret2022variational}; others adapt computation to each input, skipping layers or exiting early \cite{wang2018skipnet, huang2018multiscale, raposo2024mixture} or learning when to halt recurrence \cite{graves_adaptive_2016, banino2021pondernet} (see \cite{han2021dynamic} for a review).
In our approach, breadth and depth are learned but fixed at inference, as in brains, whereas time can be either fixed or allocated per input (Sec.~\ref{section:time}; Appendix~\ref{appendix:act} compares our time selection with ACT).
Breadth, depth, and time are known to be partly interchangeable.
For instance, shallow recurrent networks can be equivalent to deep residual networks with shared weights \cite{liao2016bridging} and can trade depth for iterations at test time \cite{schwarzschild2021can}; wide, shallow residual networks can match deeper ones \cite{zagoruyko2016wide}.
Architecture size can also be found by shrinking or growing a network.
Pruning removes the units that contribute least to performance post-hoc \cite{lecun1989optimal, han_learning_2015, li_pruning_2017, frankle_lottery_2019}, typically through iterative cycles of removal and fine-tuning, whereas constructive methods add units or layers during training \cite{fahlman1989cascade, chen2016net2net}.
We instead start from an overcomplete graph and let resource costs determine how much of it is used.
Finally, neural architecture search explores discrete architectural choices \cite{zoph2017neural, liu_progressive_2018}, including differentiable \cite{liu2018darts} and hardware-aware \cite{cai2019proxylessnas} variants, and compound scaling grows width, depth, and input resolution together at a fixed ratio found by grid search \cite{tan2019efficientnet}.
None of these approaches prices breadth, depth, and time jointly and lets the functional architecture emerge through optimization.

\section{Methods}\label{section:methods}

\subsection{Multi-resource cost (MRC) optimization}

A typical loss for a neural network includes task performance (e.g., cross-entropy) and some form of regularization (e.g., weight decay).
Our multi-resource cost (MRC) approach adds differentiable cost terms for multiple resources, which enables studying interactions between resources:

\begin{equation}
    \mathcal{L} = \underbrace{\lambda_{\text{task}}\mathcal{L}_{\text{task}}}_{\text{task}}
    \ + \
    \underbrace{
    \lambreadth\mathcal{L}_{\text{breadth}} +
    \lamdepth\mathcal{L}_{\text{depth}}}_{\text{space}} +
    \underbrace{
    \lamtime\mathcal{L}_{\text{time}}
    }_{\text{time}}
\end{equation}

The $\lambda$ coefficients control the relative price of each resource.
Higher $\lambda$ values pressure the network to use less of that resource.
We fix $\lambda_{\text{task}}=1$ throughout and vary only the price of breadth, depth, and time.
${L}_{\text{task}}$, ${L}_{\text{breadth}}$, ${L}_{\text{depth}}$, ${L}_{\text{time}}$ are defined in Sec.~\ref{subsection:costs}, after explaining the model architecture.

\subsection{Model architecture}
\label{subsection:model}

\begin{figure}
    \centering
    \includegraphics[width=\linewidth, trim={5cm 16cm 7cm 0cm}, clip]{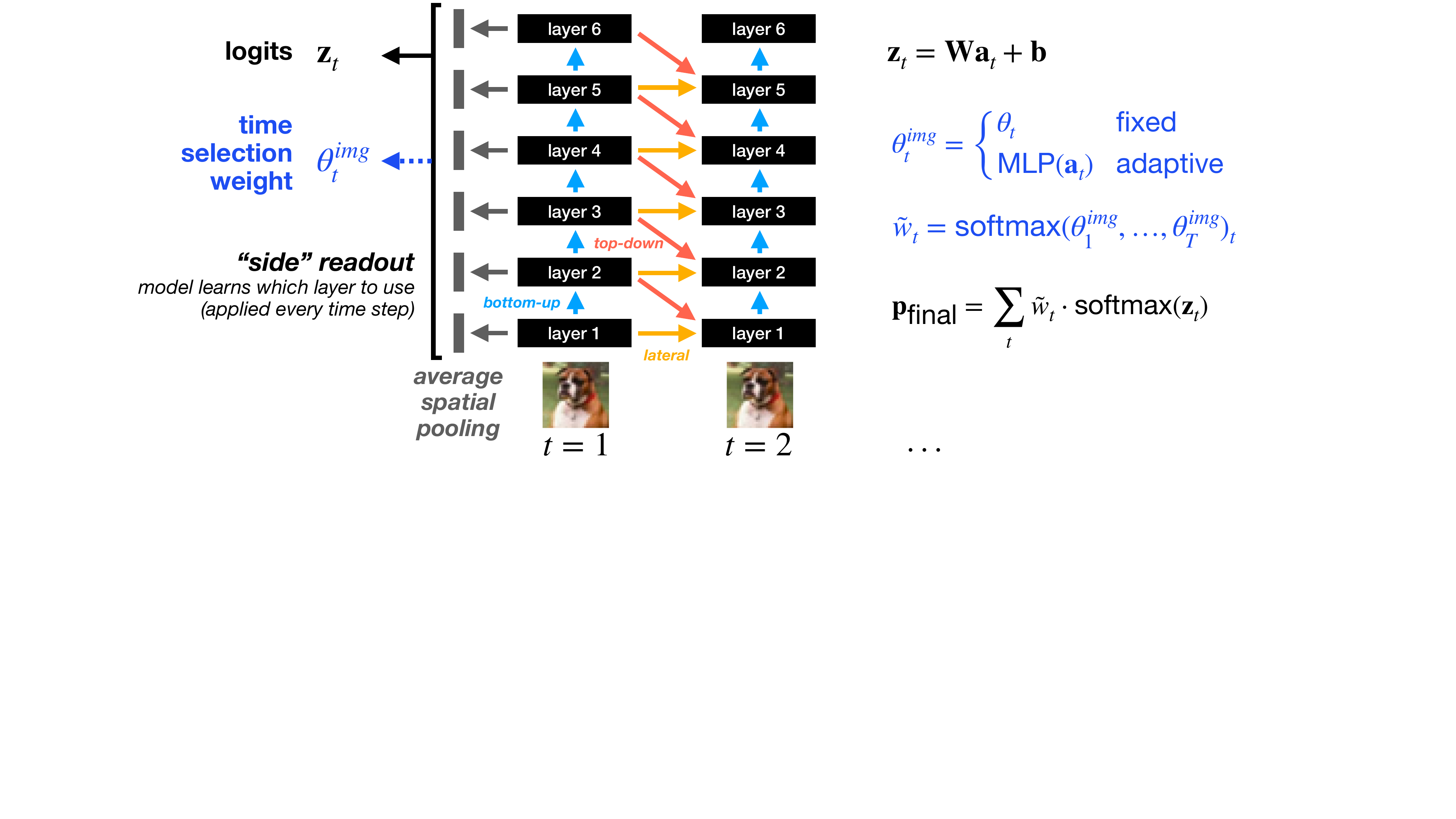}
    \caption{
    \textbf{Model architecture.}
    The network implements a finite subset of the infinite lattice (Fig.~\ref{fig:lattice-solutions}), within which computational graphs of different effective breadth (channels used), depth (layers used), and time (recurrent steps used) emerge through training.
    It is a recurrent convolutional network with bottom-up, lateral, and top-down connections ($L{=}6$, $T{=}5$, $C{=}48$, $3{\times}3$ convs), in which each layer maintains a hidden state that accumulates information over time.
    A side readout applied at every time step allows the model to use earlier layers and reduce effective depth.
    Time selection weights determine how much the final prediction relies on each time step; they are either fixed (shared across inputs) or adaptive (input-dependent via a small MLP).
}
    \label{fig:model}
\end{figure}

Our superset model architecture is meant to implement the infinite lattice in breadth, depth, and time (Fig.~\ref{fig:lattice-solutions}).
In practice, the superset architecture is implemented by a finite recurrent convolutional neural network (Fig.~\ref{fig:model}).
The idea is to grow a diverse set of functional computational graphs that use different amounts of breadth, depth, and time within this superset architecture.

\paragraph{Input.}
All images are resized to $32 \times 32 \times 3$ and passed through an initial convolution to map the input channels to $C = 48$ feature channels.

\paragraph{Recurrent convolutional network.}
The superset model is a recurrent convolutional network with $L = 6$ layers and $T = 5$ time steps.
At each time step $t$, each layer $i$ receives three inputs combined additively: a bottom-up signal from layer $i{-}1$ at the current time step, a lateral signal from layer $i$ at the previous time step, and a top-down signal from layer $i{+}1$ at the previous time step (Fig.~\ref{fig:model}).
The top layer receives no top-down input.
All connections are $3 \times 3$ convolutions with $C=48$ output channels.
Each layer maintains a hidden state $\mathbf{h}_{t,i}$ that accumulates input across time steps, followed by a ReLU nonlinearity.
Divisive normalization, a canonical computation in cortex~\cite{carandini2012normalization}, is applied to both the bottom-up input and the lateral and top-down signals.
We add Gaussian noise ($\sigma = 0.1$) to the hidden states at each time step (see motivation for noise in Appendix \ref{appendix:noise}).

\paragraph{Side readout.}
At each time step $t$, the post-ReLU hidden states are spatially average-pooled and concatenated across layers, yielding $\mathbf{a}_t \in \mathbb{R}^{L \cdot C}$, to which we also add Gaussian noise ($\sigma = 0.1$; Appendix~\ref{appendix:noise}).
A linear layer maps this to class logits: $\mathbf{z}_t = \mathbf{W}\mathbf{a}_t + \mathbf{b}$.
The side readout mechanism allows the model to decide how many layers to use.

\paragraph{Time selection.}
The model produces class logits at every time step and time selection weights determine how those logits are combined into a single prediction.
We consider two schemes for time selection.
\textit{Fixed} time selection uses learnable parameters $\theta_1, \dots, \theta_T$ shared across all inputs; the model learns a single temporal allocation applied uniformly.
\textit{Adaptive} time selection computes a scalar weight at each time step as a function of the current pooled image-dependent activations via a small two-layer MLP: $\theta^{\text{img}}_t = \text{MLP}(\mathbf{a}_t)$.
This allows the model to dynamically allocate different amounts of processing time to different inputs.
In both cases, the raw time selection weights are $\ell_2$-normalized, multiplied by a fixed inverse temperature ($\tau = 4$), and passed through a softmax to obtain the time selection weights $\tilde{w}_1, \dots, \tilde{w}_T$ ($\tilde{w}_t \in [0.003, 0.956]$, ensuring gradients flow to all time steps during training).
See Appendix~\ref{appendix:act} for comparison to ACT \cite{graves_adaptive_2016}.

\paragraph{Output.}
The final output is a mixture of per-timestep probability distributions, weighted by the time selection weights:
$\mathbf{p}_{\text{final}} = \sum_t \tilde{w}_t \cdot \text{softmax}(\mathbf{z}_t)$.

\subsection{Costs}
\label{subsection:costs}

\paragraph{Task.}
The task cost is the cross-entropy between model predictions and dataset labels:

\begin{equation}
    \mathcal{L}_{\text{task}} = -\frac{1}{\log K} \frac{1}{N} \sum_{n=1}^{N} \log p_{n,y_n}
    \label{eq:task}
\end{equation}

where $p_{n,y_n}$ is the predicted probability for the correct class of sample $n$ and $K$ is the number of classes.
Normalization by $\log K$ ensures the error cost is comparable across datasets with different numbers of classes (e.g., $K=10$ for CIFAR-10 vs.\ $K=200$ for Tiny ImageNet), with a value of 1.0 corresponding to chance-level performance.

\paragraph{Breadth.}
Within each layer, output channels are sorted by their magnitude $\bar{w}$ (defined below).
The breadth cost applies a weight decay that scales with channel rank:
\begin{equation}
    \mathcal{L}_{\text{breadth}} = \frac{1}{LC} \sum_{i=1}^{L} \sum_{k=1}^{C} \beta_k \, \bar{w}_{i,\pi_i(k)},
    \qquad
    \beta_k = \frac{(k/C)^{\gamma}}{\frac{1}{C}\sum_{j=1}^{C} (j/C)^{\gamma}}
    \label{eq:breadth}
\end{equation}
where $\bar{w}_{i,\pi_i(k)}$ is the magnitude of the channel with rank $k$ in layer $i$: the mean absolute weight of each kernel (bottom-up, lateral, and top-down; the top layer has no top-down input) feeding that channel, summed over the kernel types. $\pi_i$ is the permutation that sorts channels by descending magnitude, and $\gamma$ is a position power that controls how steep the penalty is ($\gamma=4$, chosen to produce a sharp cutoff between used and unused channels).
The position weights $\beta_k$ are normalized to have mean 1, so the overall scale of the cost does not depend on $\gamma$.
Because the channel ranking is held fixed during backpropagation, the breadth cost is differentiable (Appendix~\ref{appendix:differentiability}).
The rank-dependent scaling allows the network to organically consolidate useful features into a few high-magnitude channels while pushing unused channels toward zero.

\paragraph{Depth.}
The depth cost applies a weight decay that scales with layer index:
\begin{equation}
    \mathcal{L}_{\text{depth}} = \frac{1}{LC} \sum_{i=1}^{L} \delta_i \sum_{k=1}^{C} \bar{w}_{i,k},
    \qquad
    \delta_i = \frac{(i/L)^{\gamma}}{\frac{1}{L}\sum_{j=1}^{L} (j/L)^{\gamma}}
\end{equation}
The depth cost pressures the network to solve the task with fewer layers when possible.
The same position power ($\gamma = 4$) and mean-1 normalization are applied as for the breadth cost.

\paragraph{Time.}
The time cost is the expected time step under the time selection weights, scaled by $1/(T-1)$:
\begin{equation}
    \mathcal{L}_{\text{time}} = \sum_{t=1}^{T} \tilde{w}_t \cdot \frac{t}{T-1}
\end{equation}
where $\tilde{w}_t$ are the time selection weights.
For \textit{fixed} time selection, this cost is the same for all inputs.
For \textit{adaptive} time selection, the cost varies per input---the network learns to spend more time on images where the reduction in error cost outweighs the additional time cost.

\paragraph{Optimization.}
Resource costs and noise are both annealed during training to ensure stable learning (Appendix~\ref{appendix:annealing}).
All models are trained for 150 epochs using AdamW with cosine learning rate decay (Appendix~\ref{appendix:training-details}).
Each experimental condition is trained across multiple independent instances with different random initializations (Appendix~\ref{appendix:experiments}).

\subsection{Definitions of resources used}

To quantify the effective resources used by each trained model, we apply a post-hoc pruning procedure that identifies the smallest sub-network preserving 98\% of above-chance accuracy (details in Appendix~\ref{appendix:pruning}; robustness to different accuracy thresholds in Appendix~\ref{appendix:pruning-threshold}).
The result of the pruning procedure is a binary mask over layers and channels, from which we define:

\textbf{Layers used (depth):} the number of layers with at least one surviving channel.

\textbf{Channels used (breadth):} the average number of surviving channels per active layer.

\textbf{Time used:} the expected time step index under the time selection weights, $\sum_t \tilde{w}_t \cdot t$.

\subsection{Experiments}

We use CIFAR-10 \cite{krizhevsky2009learning} as our main dataset across all experiments.
Full experimental configurations are provided in Appendix~\ref{appendix:experiments}.
Error bars denote 95\% confidence intervals across model instances using the $t$-distribution. Shaded regions in Fig.~\ref{fig:time}c,g,h denote 95\% bootstrap confidence intervals.

\paragraph{Breadth vs.\ depth.}
We vary $\lambreadth$ and $\lamdepth$ over $\{0, 1, 10, 10^2, 10^3, 10^4\}$ each with $\lamtime = 0$, yielding a $6 \times 6$ grid of resource pressure combinations.

\paragraph{Time.}
We vary $\lamtime$ from 0 to 1 in increments of 0.1, comparing fixed and adaptive time selection schemes  (no space costs: $\lambreadth = \lamdepth = 0$).
This is the only experiment that has model instances with fixed time selection---all other experiments use the adaptive time selection scheme.

\paragraph{Breadth vs.\ depth vs.\ time.}
We vary all three cost factors jointly, combining the breadth and depth grid above with six levels of $\lamtime \in \{0, 0.1, 0.2, 0.3, 0.5, 1.0\}$.

\paragraph{Task complexity.}
We compare MNIST \cite{lecun1998gradient}, CIFAR-10, and Tiny ImageNet \cite{le2015tiny} (a 200-class subset of ImageNet \cite{deng2009imagenet}), varying $\lambreadth$, $\lamdepth$, and $\lamtime$ identically across all three datasets to test whether networks grow with task complexity (see operationalization in Appendix \ref{appendix:task-complexity}).

\section{Results}\label{section:results}

\subsection{Breadth vs.\ depth}

\begin{figure}[t]
    \centering
    \resizebox{0.95\textwidth}{!}{%
    \begin{tikzpicture}
        % Define row y-positions (adjust as needed)
        \def\rowone{0}
        \def\rowtwo{-2.5cm}
        \def\rowthree{-5.5cm}
        \def\rowfour{-8.6cm}

        % Row 1: a (costs) left, b (norms) right
        \node[anchor=north west, inner sep=0] (costs) at (0, \rowone)
            {\includegraphics[width=0.35\linewidth]{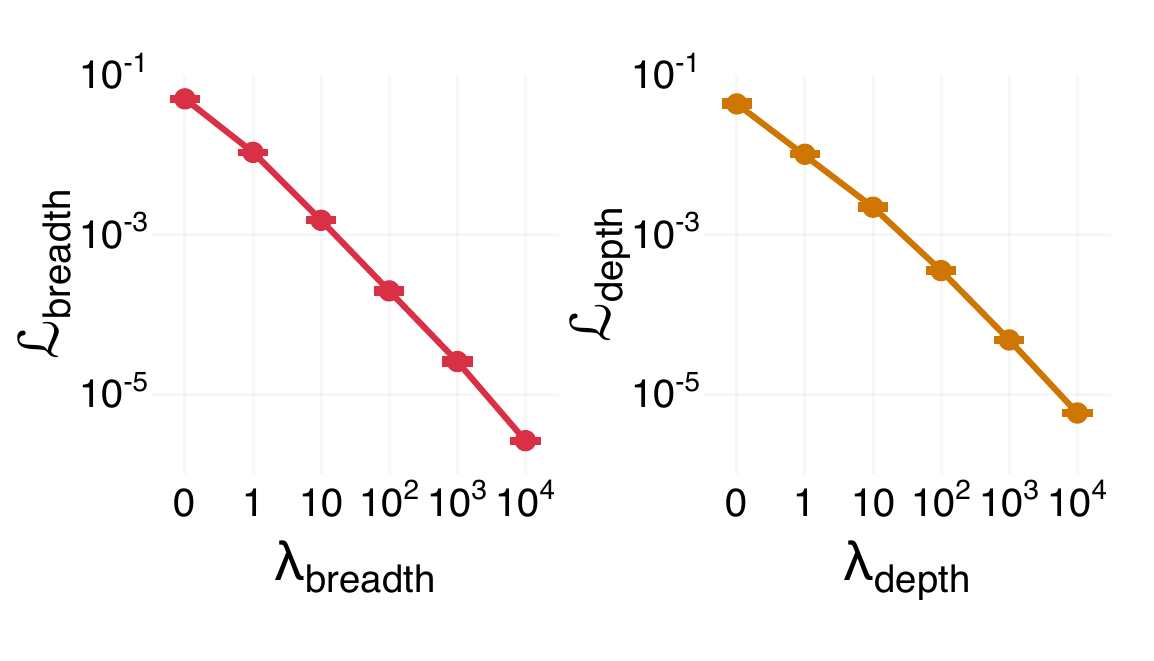}};
        \node[anchor=north west, inner sep=0] (norms) at (0.37\linewidth, \rowone)
            {\includegraphics[trim={-0.1cm 0cm 0 4.5cm}, clip, width=0.61\linewidth]{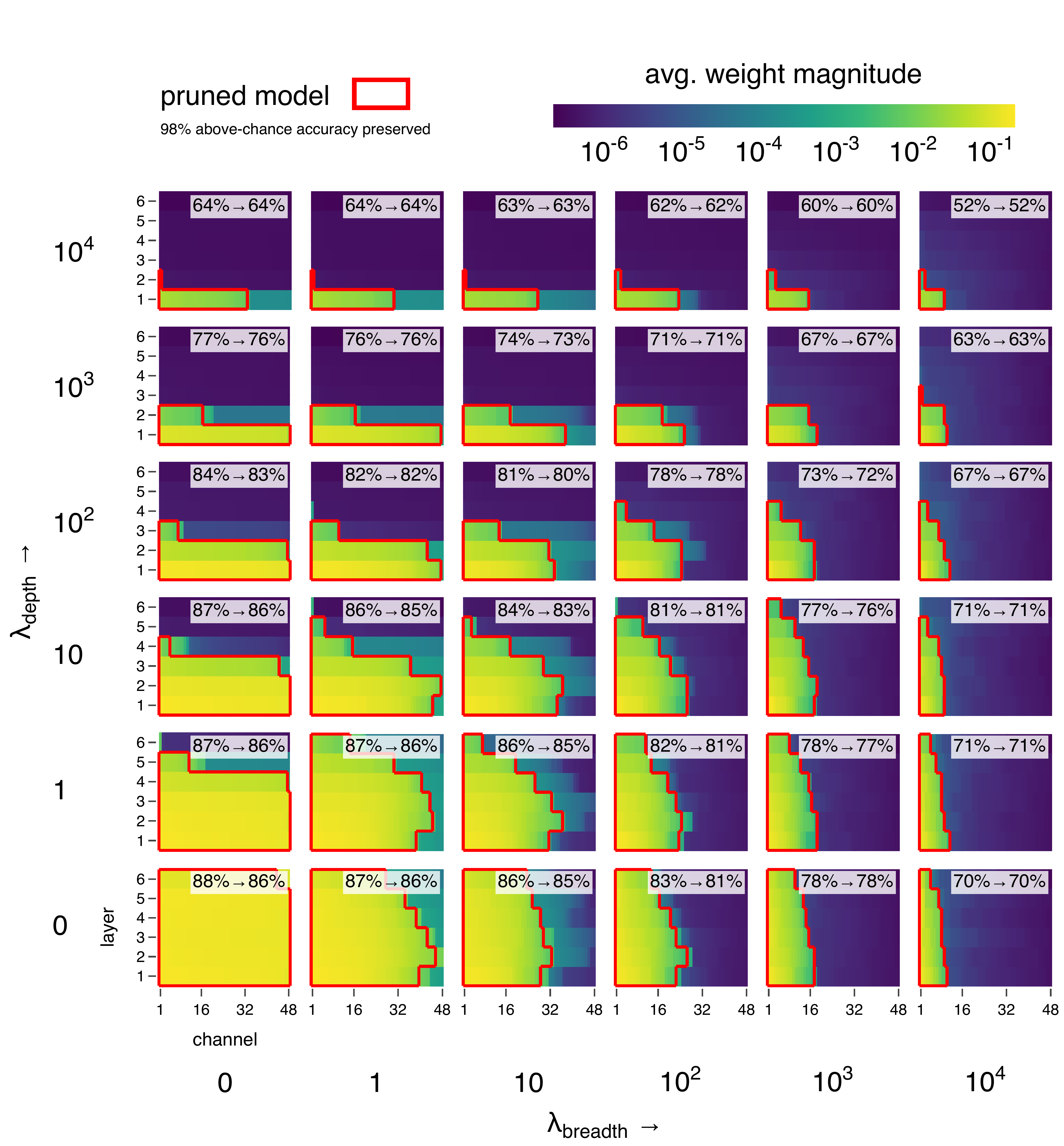}};

        % Row 2: c (resources) left
        \node[anchor=north west, inner sep=0] (resources) at (0, \rowtwo)
            {\includegraphics[width=0.35\linewidth]{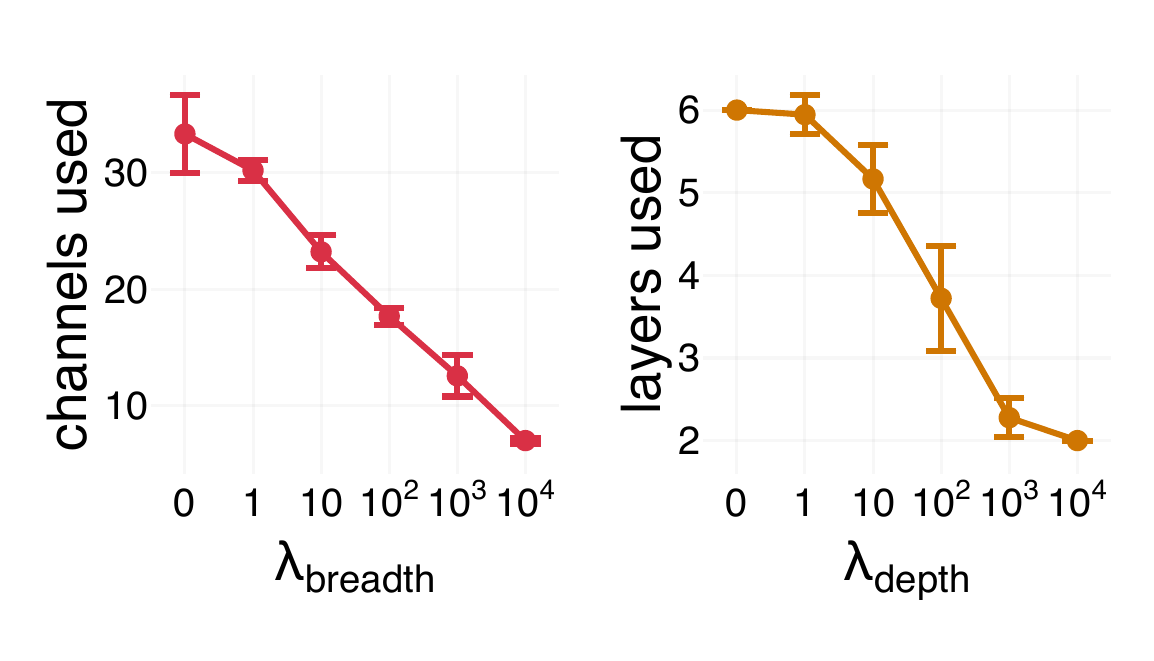}};

        % Row 3: d (accuracy) left
        \node[anchor=north west, inner sep=0] (accuracy) at (0, \rowthree)
            {\includegraphics[trim={-1cm 0 0 0}, clip, width=0.35\linewidth]{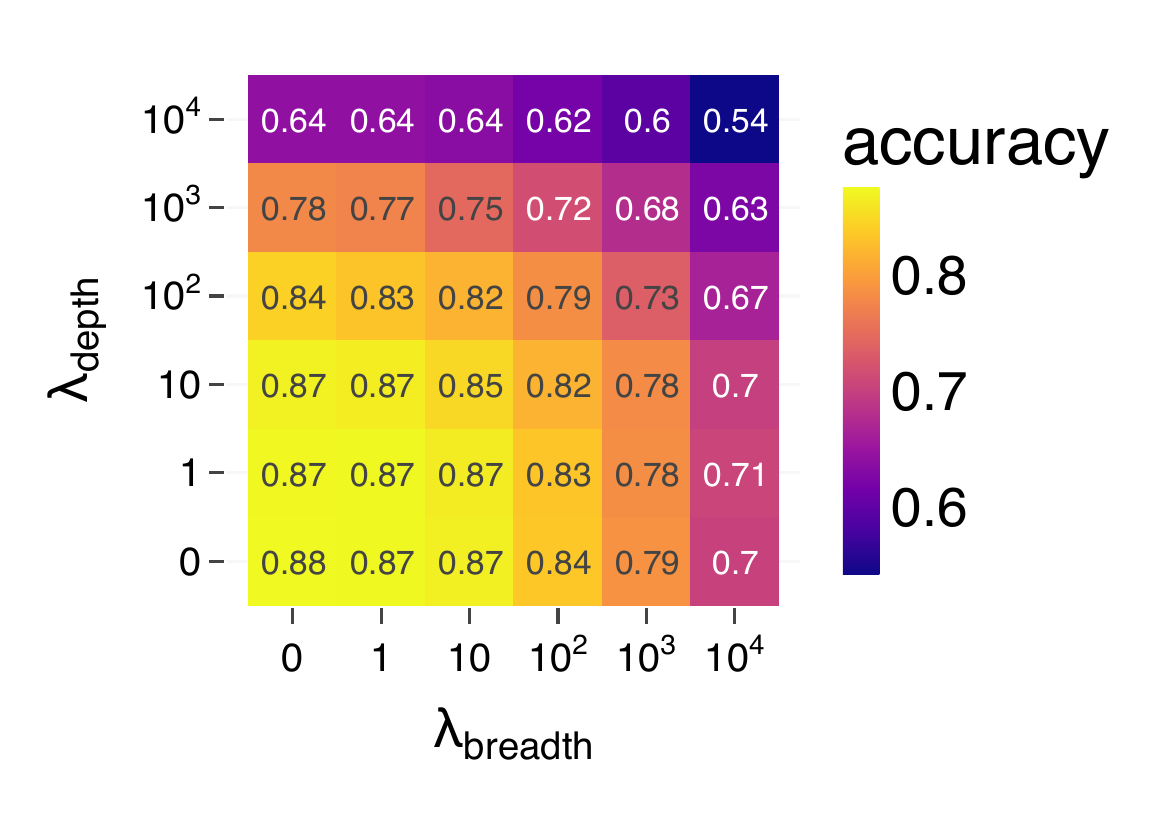}};

        % Row 4: e (attribution) left, f (entropy) right
        \node[anchor=north west, inner sep=0] (attribution) at (0, \rowfour)
            {\includegraphics[trim={0cm 0cm 0 0cm}, clip, width=0.67\linewidth]{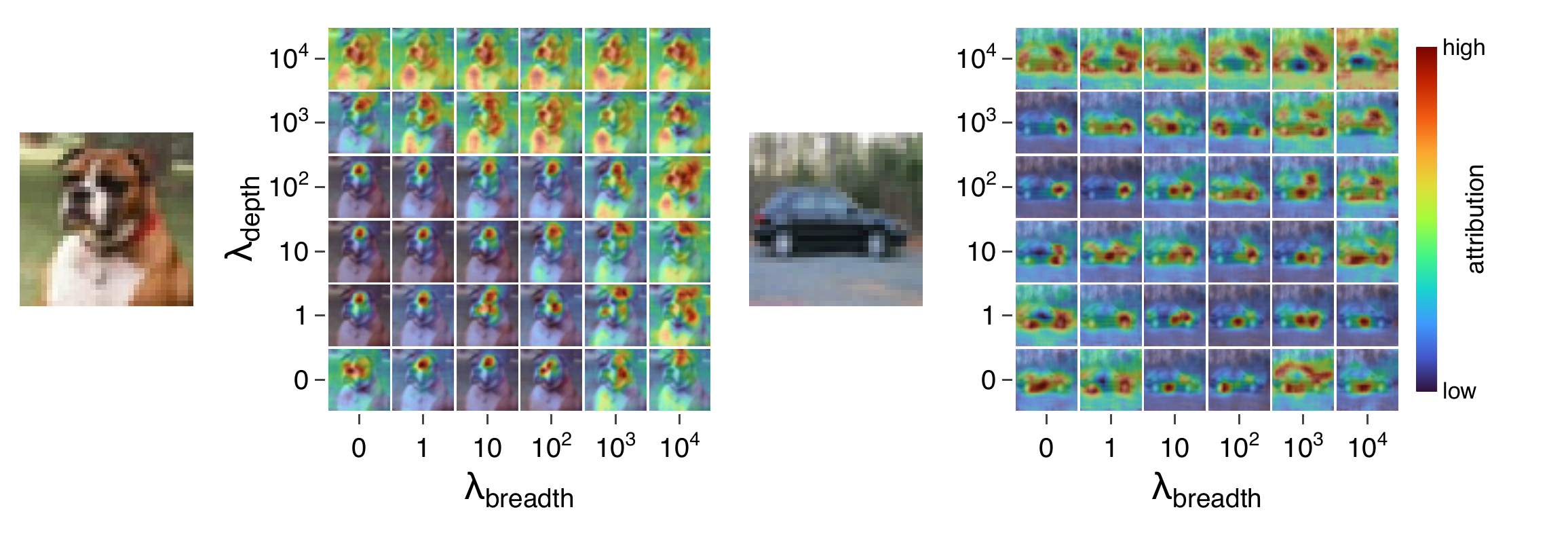}};
        \node[anchor=north west, inner sep=0] (attentropy) at (0.7\linewidth, \rowfour)
            {\includegraphics[width=0.3\linewidth]{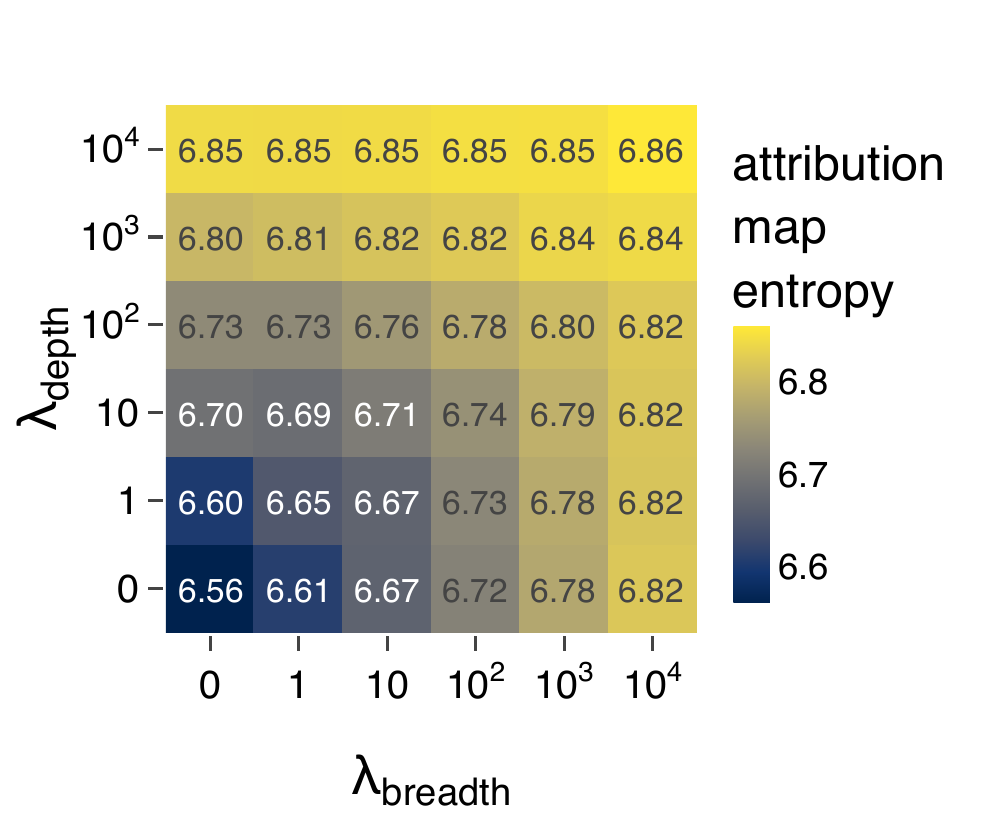}};

        % Panel labels
        \node[font=\bfseries] at ([shift={(-0.0cm,-0.1cm)}]costs.north west) {a};
        \node[font=\bfseries] at ([shift={(-0.0cm,-0.1cm)}]norms.north west) {b};
        \node[font=\bfseries] at ([shift={(-0.0cm,-0.1cm)}]resources.north west) {c};
        \node[font=\bfseries] at ([shift={(-0.0cm,-0.1cm)}]accuracy.north west) {d};
        \node[font=\bfseries] at ([shift={(-0.0cm,-0.3cm)}]attribution.north west) {e};
        \node[font=\bfseries] at ([shift={(-0.0cm,-0.3cm)}]attentropy.north west) {f};

    \end{tikzpicture}
    }
    \caption{
    \textbf{Breadth vs. depth.}
    ~\textbf{a}~
    Raw costs decrease smoothly with increasing $\lambreadth$ and $\lamdepth$.
    ~\textbf{b}~
     Average weight magnitudes across layers and channels (one model instance per $\lambda$ combination), with pruned-model boundaries in red (98\% of above-chance accuracy preserved) and accuracy before $\rightarrow$ after pruning (top right).
    ~\textbf{c}~
    Pruning-defined resources used (channels, layers) decrease as a function of $\lambda$.
    ~\textbf{d}~
    Accuracy decreases with breadth and depth pressure.
    ~\textbf{e}~
    Attribution maps using input perturbation for two example images (Appendix~\ref{appendix:attribution}).
    Constrained models draw on evidence spread across the image, while less constrained models rely on localized object parts (e.g., the dog's face).
    ~\textbf{f}~
    Attribution map entropy (across 103 images and 3 model instances) quantifying attribution map spread as a function of $\lambreadth$ and $\lamdepth$.
    }
    \label{fig:breadth-depth}
\end{figure}

We begin by considering breadth and depth costs alone, setting $\lamtime = 0$.
Raw costs $\mathcal{L}_{\text{breadth}}$ and $\mathcal{L}_{\text{depth}}$ decrease smoothly with increasing $\lambreadth$ and $\lamdepth$ (Fig.~\ref{fig:breadth-depth}a), confirming that the differentiable cost terms work as intended.

To understand the solutions that emerge, we visualize the average weight magnitude across layers and channels for each $\lambda$ combination (Fig.~\ref{fig:breadth-depth}b).
As costs increase, weights concentrate into fewer layers and channels, with the pruning boundary (red outline) shrinking accordingly.
Our pruning procedure (Appendix~\ref{appendix:pruning}) recovers compact sub-networks that preserve 98\% of above-chance accuracy without fine-tuning, confirming that the resource costs produce genuinely sparse solutions.
The pruning-defined resources, channels used and layers used, decrease as a function of $\lambda$ (Fig.~\ref{fig:breadth-depth}c).

Critically, breadth and depth can be traded off at a given level of performance (Fig.~\ref{fig:breadth-depth}b,d).
Shallow-and-wide models (high $\lamdepth$, low $\lambreadth$) can achieve accuracy comparable to narrow-and-deep models (low $\lamdepth$, high $\lambreadth$).

Finally, we ask whether models with different resource profiles rely on different features.
Inspired by the texture bias in ImageNet-trained CNNs~\cite{geirhos2019imagenet}, we hypothesized that shallow models, with fewer stages to build up object-level representations, would draw on evidence spread across the image.
Attribution maps from input perturbation \cite{zeiler2014visualizing} are consistent with this: heavily depth-constrained models rely on features spread across the image, whereas less constrained models focus on localized and interpretable object parts such as the dog's face (Fig.~\ref{fig:breadth-depth}e).
However, attribution map entropy increases with breadth pressure as well (Fig.~\ref{fig:breadth-depth}f).
At matched accuracy, shallower models show a small trend toward higher entropy (Fig.~\ref{fig:entropy-vs-accuracy}); a more controlled investigation is left for future work.

\subsection{Time: emergent adaptive processing}
\label{section:time}

\begin{figure}[t]
    \centering
    \resizebox{0.93\textwidth}{!}{%
    \begin{tikzpicture}
        % Define row y-positions (adjust these as needed)
        \def\rowone{0}
        \def\rowtwo{-3.5cm}
        \def\rowthree{-6.6cm}

        % Row 1: a b c d
        \node[anchor=north west, inner sep=0] (cost) at (0, \rowone)
            {\includegraphics[width=0.22\linewidth]{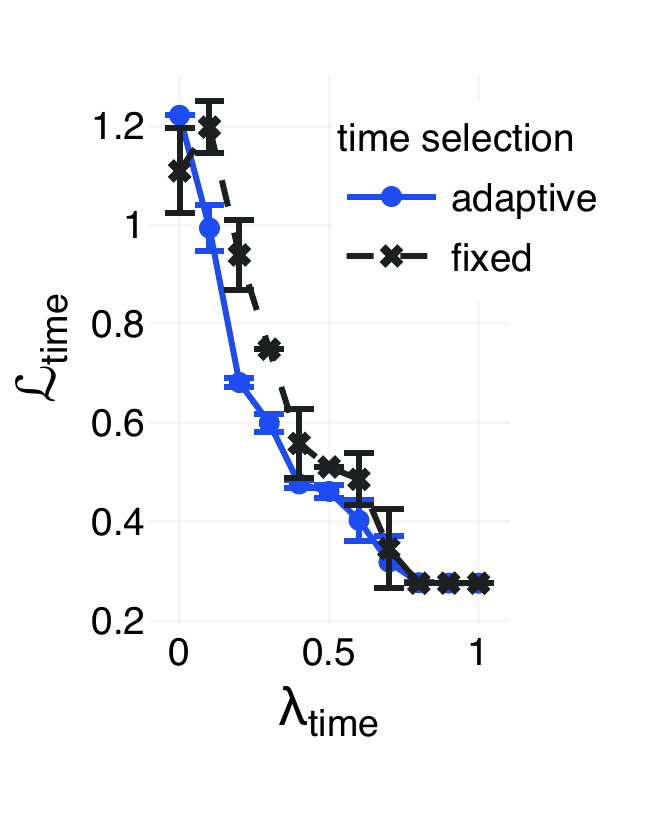}};
        \node[anchor=north west, inner sep=0] (timeused) at (0.22\linewidth, \rowone)
            {\includegraphics[width=0.22\linewidth]{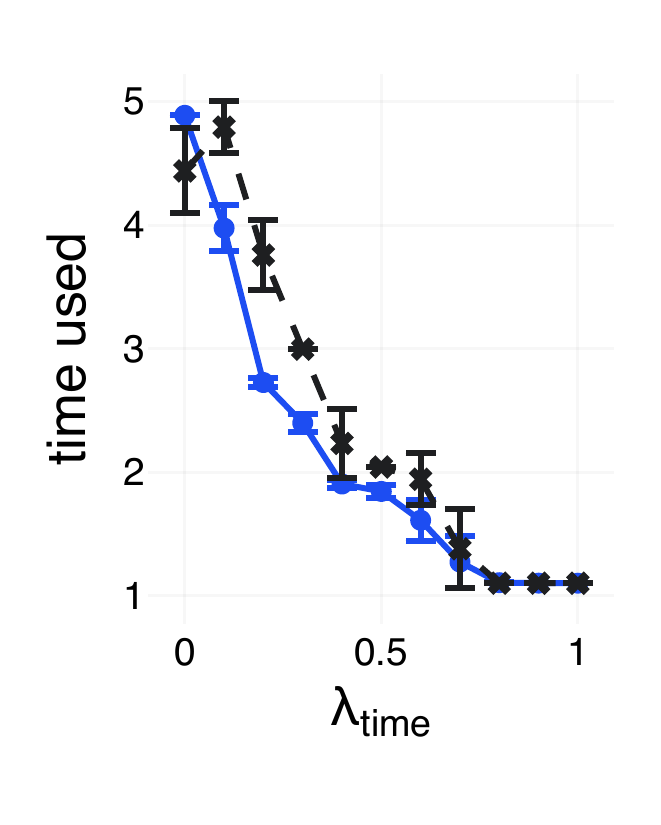}};
        \node[anchor=north west, inner sep=0] (acc) at (0.45\linewidth, \rowone)
            {\includegraphics[width=0.27\linewidth]{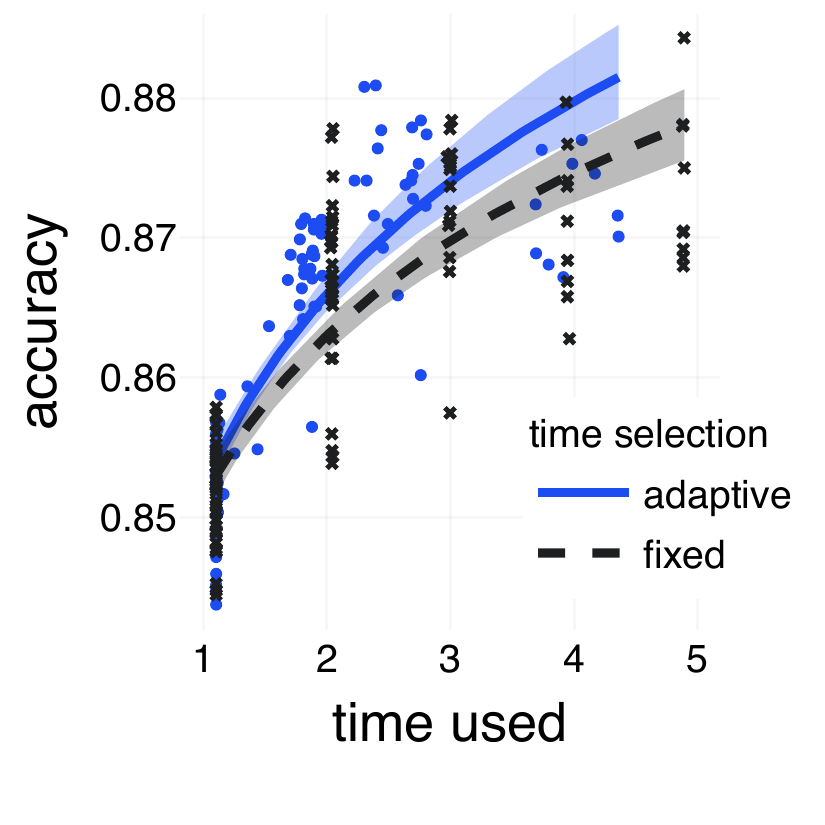}};
        \node[anchor=north west, inner sep=0] (occlusion) at (0.72\linewidth, \rowone)
            {\includegraphics[width=0.27\linewidth]{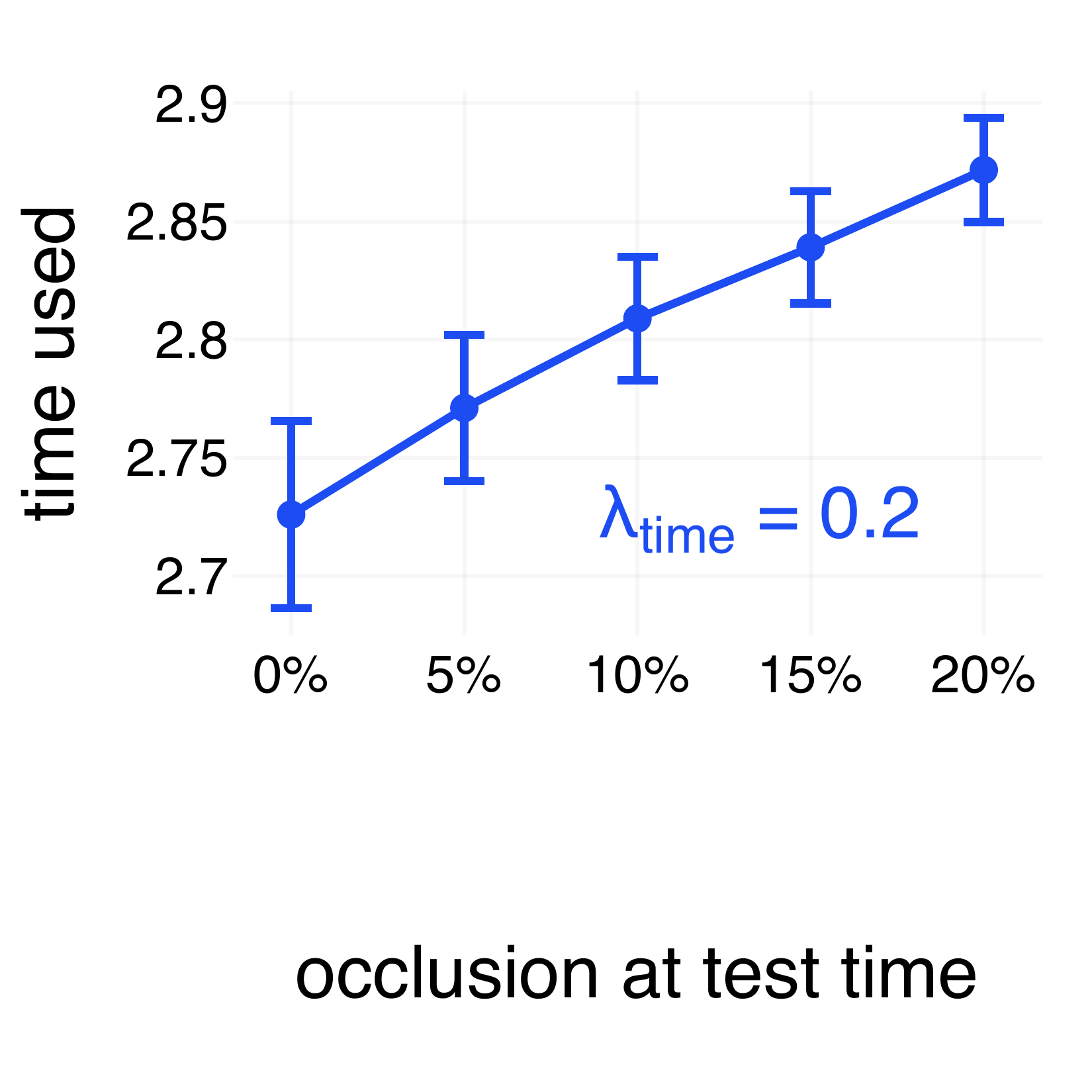}};
            
        % Occlusion examples inset
        \node[anchor=south west, inner sep=0] (occ_examples) at ([xshift=0.65cm, yshift=0.6cm]occlusion.south west)
            {\includegraphics[width=0.22\linewidth]{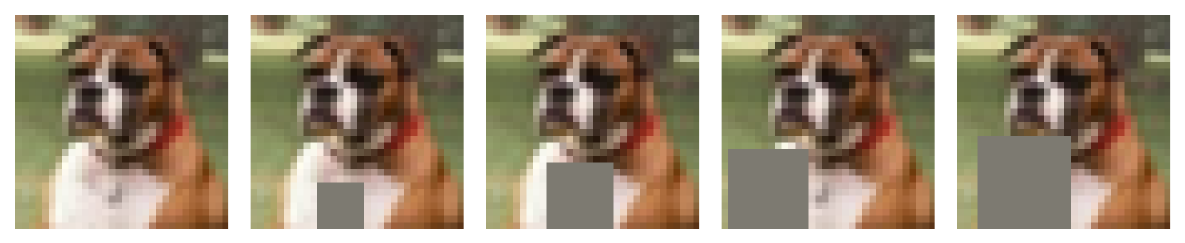}};

        % Row 2: e f
        \node[anchor=north west, inner sep=0] (images) at (0, \rowtwo)
            {\includegraphics[width=0.48\linewidth]{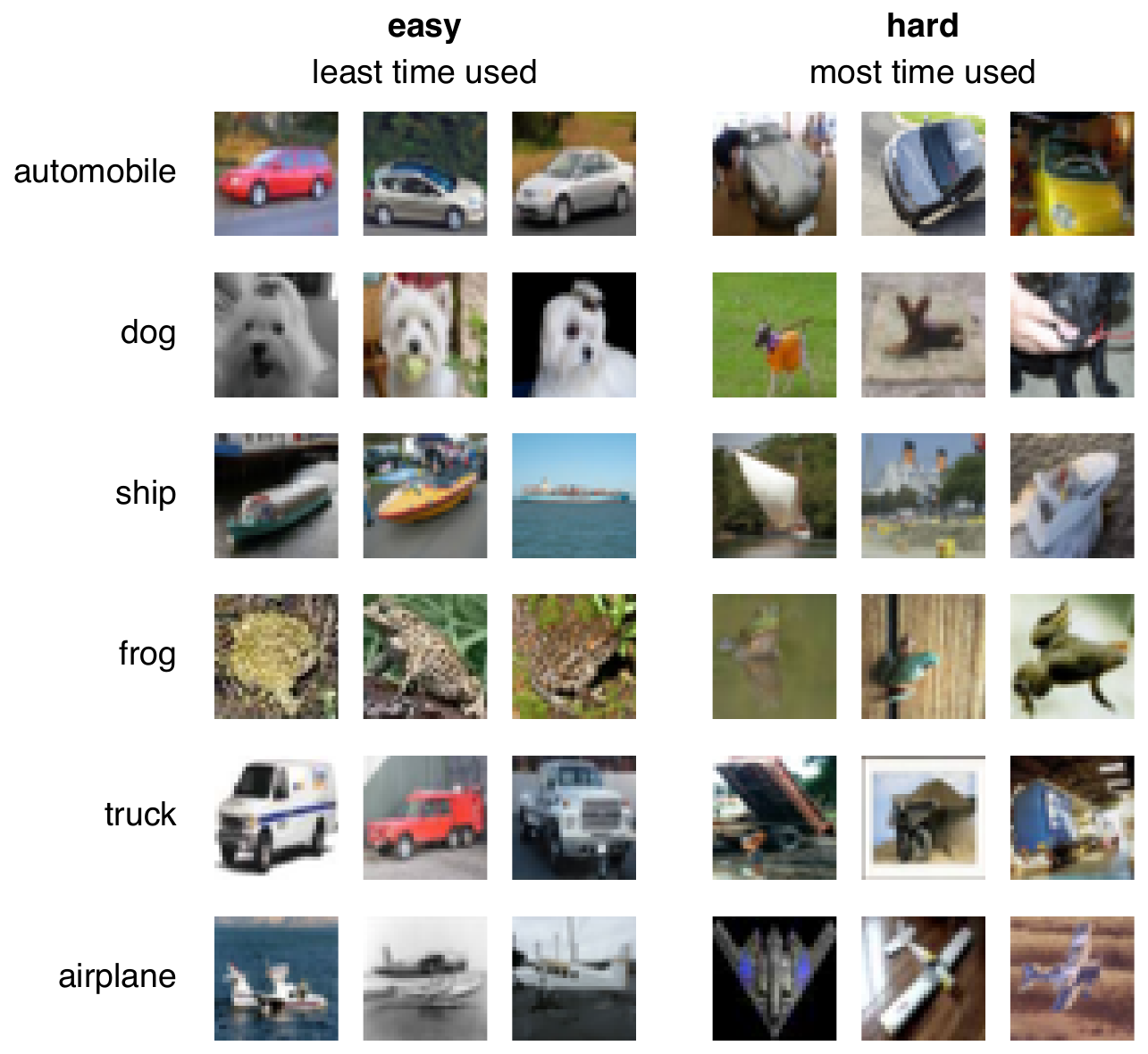}};
        \node[anchor=north west, inner sep=0] (category) at (0.55\linewidth, \rowtwo)
            {\includegraphics[width=0.40\linewidth]{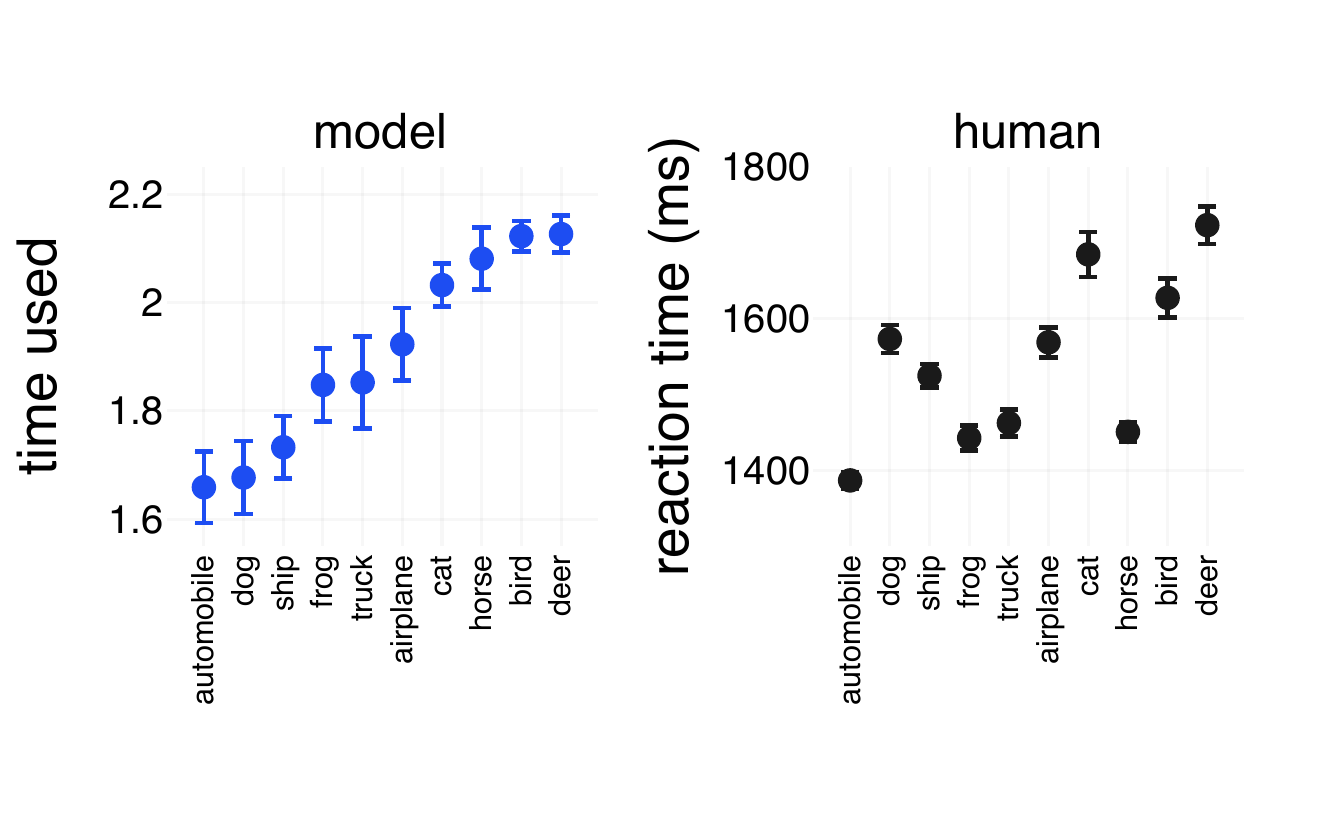}};

        % Row 3: g h
        \node[anchor=north west, inner sep=0] (correlationrt) at (0.55\linewidth, \rowthree)
            {\includegraphics[width=0.18\linewidth]{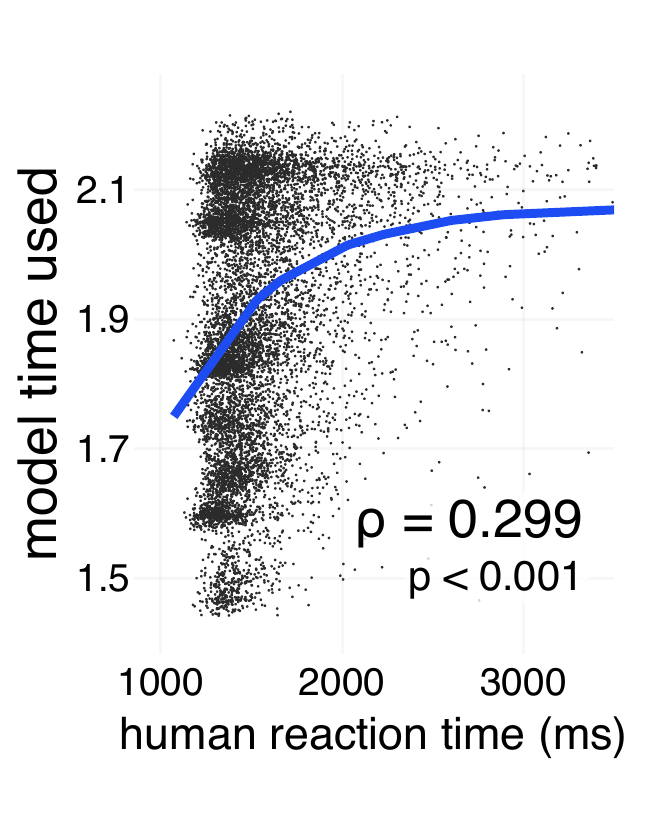}};
        \node[anchor=north west, inner sep=0] (correlation) at (0.78\linewidth, \rowthree)
            {\includegraphics[width=0.18\linewidth]{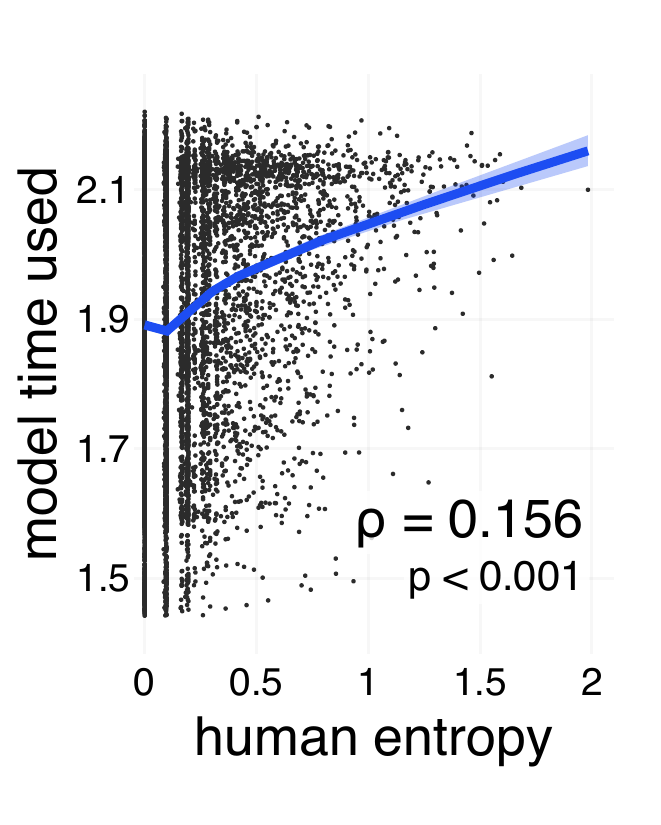}};

        % Panel labels
        \node[font=\bfseries] at ([shift={(-0.cm,-0.1cm)}]cost.north west) {a};
        \node[font=\bfseries] at ([shift={(-0.cm,-0.1cm)}]timeused.north west) {b};
        \node[font=\bfseries] at ([shift={(-0.cm,-0.1cm)}]acc.north west) {c};
        \node[font=\bfseries] at ([shift={(-0.cm,-0.1cm)}]occlusion.north west) {d};
        \node[font=\bfseries] at ([shift={(-0.cm,-0.3cm)}]images.north west) {e};
        \node[font=\bfseries] at ([shift={(-0.cm,-0.3cm)}]category.north west) {f};
        \node[font=\bfseries] at ([shift={(-0.cm,-0.4cm)}]correlationrt.north west) {g};
        \node[font=\bfseries] at ([shift={(-0.cm,-0.4cm)}]correlation.north west) {h};
    \end{tikzpicture}
    }
    \caption{
    \textbf{Time.}
    ~\textbf{a}~$\lambda_{\text{time}}$ vs.\ time cost $\mathcal{L}_{\text{time}}$.
    ~\textbf{b}~$\lambda_{\text{time}}$ vs.\ time used.
    ~\textbf{c}~Time used vs.\ accuracy: adaptive time selection dominates fixed.
    ~\textbf{d}~Occlusion introduced at test time increases time used, demonstrating that the model adaptively chooses how long to compute (Appendix~\ref{appendix:occlusion-experiment}).
    ~\textbf{e}--\textbf{h}~Adaptive model behavior averaged across all $\lamtime > 0$.
    ~\textbf{e}~Easy and hard images for several categories, defined by model time used.
    Model spends more time on ambiguous images.
    ~\textbf{f}~Average model time used and human reaction time per category, sorted by model time used.
    ~\textbf{g}~Image-level correlation between median human reaction times and model time used.
    ~\textbf{h}~Image-level correlation between human judgment uncertainty (entropy across participants) and model time used.
    Each point is one CIFAR-10 test image; the blue line is a LOWESS fit.
    }
    \label{fig:time}
\end{figure}

\begin{figure}[t]
    \centering
    \resizebox{0.95\textwidth}{!}{%
     \begin{tikzpicture}
        % Panel a
        \node[anchor=north west, inner sep=0] (a) at (0, 0)
            {\includegraphics[width=\textwidth]{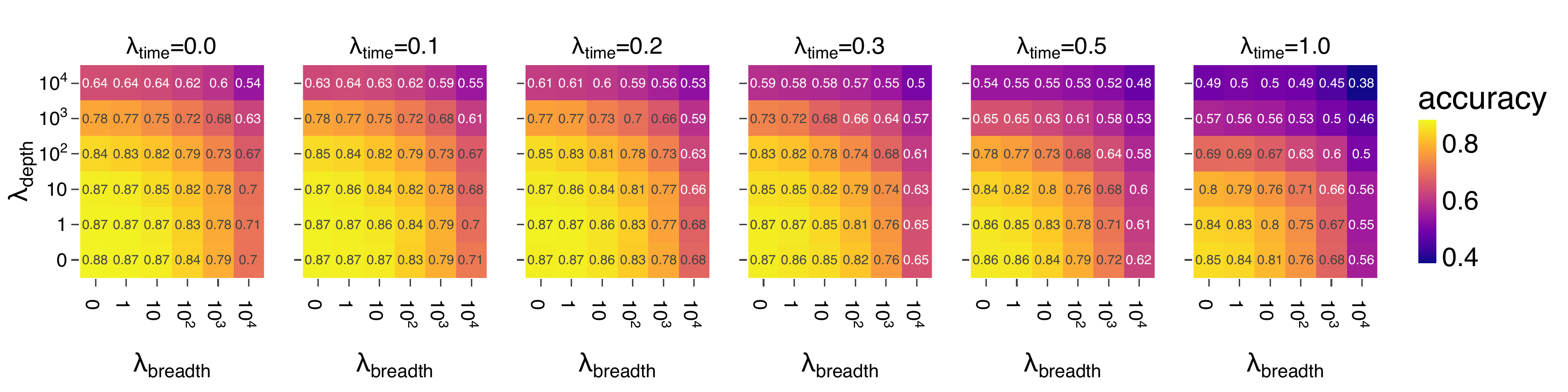}};
        
        % Panel b
        \node[anchor=north west, inner sep=0] (b) at (0, -3.5)
            {\includegraphics[width=0.26\textwidth]{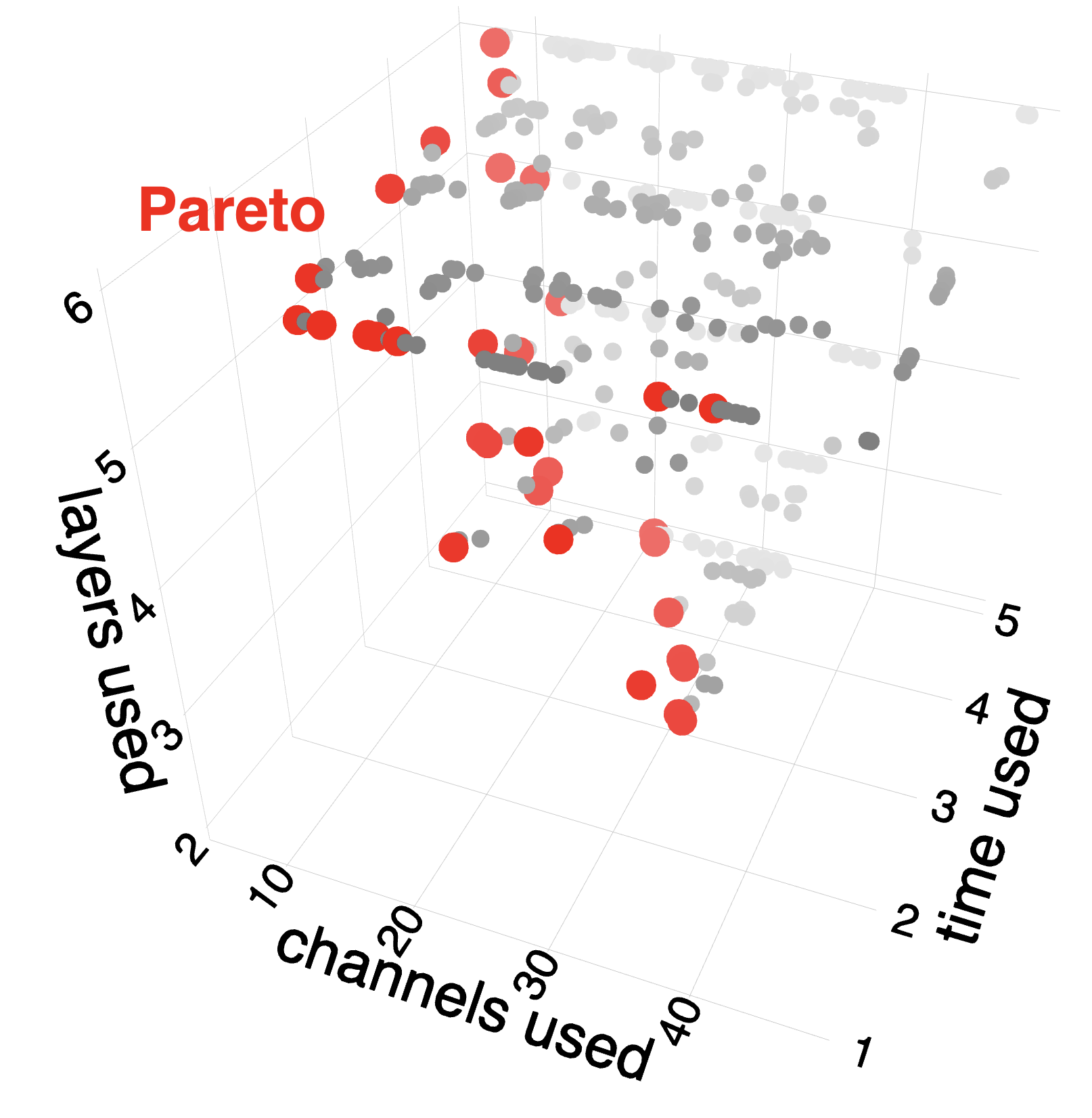}};
        
        % Panel c
        \node[anchor=north west, inner sep=0] (c) at (0.35\textwidth, -3.5)
            {\includegraphics[width=0.6\textwidth]{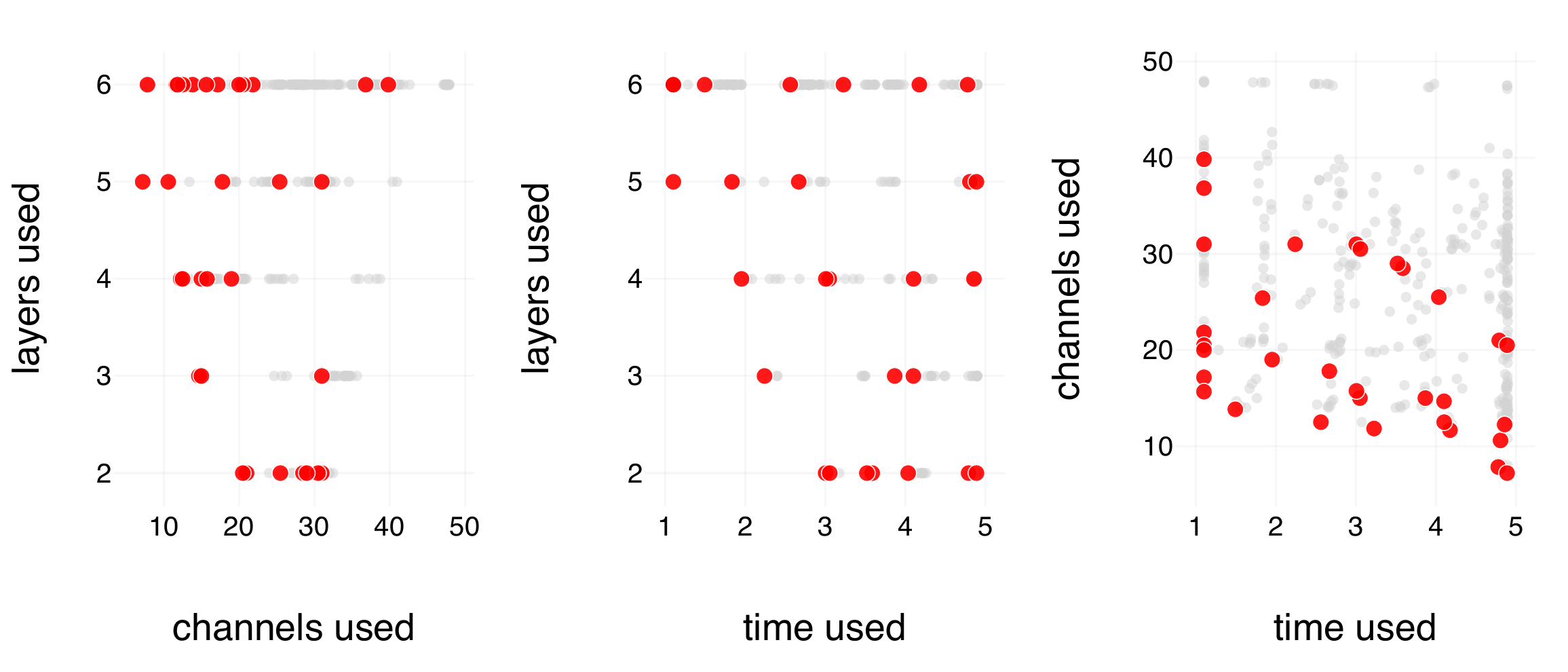}};
            
        % Panel d
        \node[anchor=north west, inner sep=0] (d) at (0, -7.1)
        {\includegraphics[width=\textwidth]{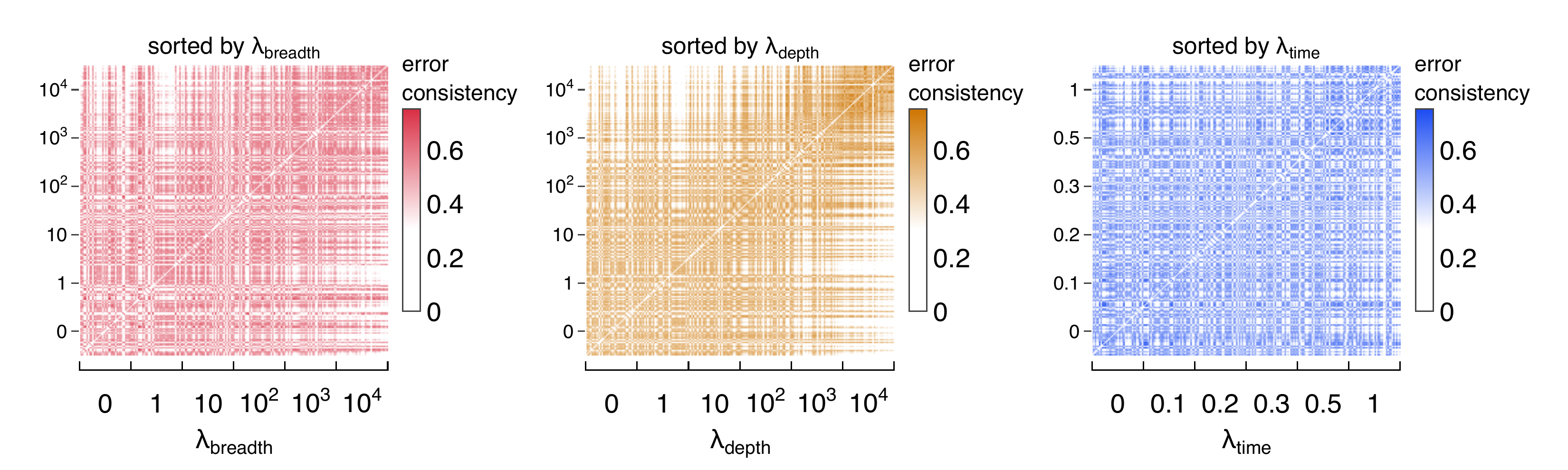}};

        % Panel e
        \node[anchor=north west, inner sep=0] (e) at (0, -11.1)
        {\includegraphics[width=\textwidth]{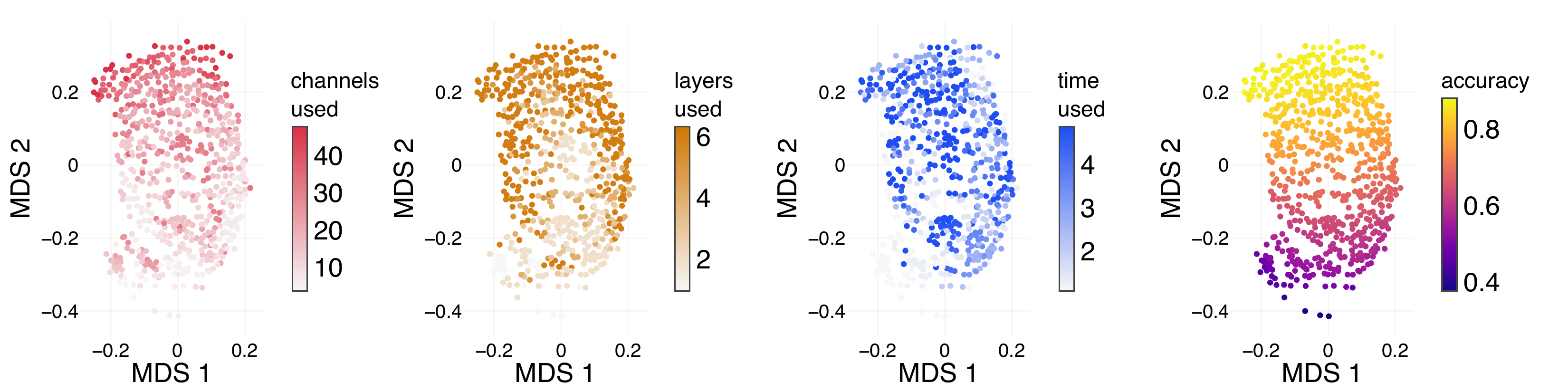}};

        % Labels
        \node[font=\normalsize\bfseries] at (a.north west) [anchor=north west, xshift=0pt, yshift=0pt] {a};
        \node[font=\normalsize\bfseries] at (b.north west) [anchor=north west, xshift=0pt, yshift=0pt] {b};
        \node[font=\normalsize\bfseries] at (c.north west) [anchor=north west, xshift=-10pt, yshift=0pt] {c};
        \node[font=\normalsize\bfseries] at (d.north west) [anchor=north west, xshift=0pt, yshift=0pt] {d};
        \node[font=\normalsize\bfseries] at (e.north west) [anchor=north west, xshift=0pt, yshift=0pt] {e};
    \end{tikzpicture}
    }
    \caption{
    \textbf{Breadth vs.\ depth vs.\ time.}
    ~\textbf{a}~Accuracy as a function of $\lambreadth$ and $\lamdepth$ for increasing $\lamtime$ (left to right).
    ~\textbf{b}~
    Empirical Pareto set (red) in resource use among models with $\geq$ 70\% accuracy.
    ~\textbf{c}~Pairwise 2D projections of the Pareto set. Red points spread across all projections, indicating that breadth, depth, and time are fungible.
    ~\textbf{d}~Error consistency between model configurations (chance-corrected), sorted by each resource cost. Depth sorting reveals a block structure: shallow and deep models tend to make different sets of errors.
    ~\textbf{e}~MDS embedding based on pairwise Jensen-Shannon divergence of output distributions, colored by channels used, layers used, time used, and accuracy.
    }
    \label{fig:breadth-depth-time}
\end{figure}

We now turn to time, setting $\lambreadth = \lamdepth = 0$.
In our setup, breadth and depth are fixed properties of the architecture at inference, whereas time can be allocated per input, i.e., the network can rely more on later steps for some inputs than for others.
We therefore compare two schemes: \textit{fixed} time selection treats time like space, fixing the time selection weights and using the same allocation for all inputs, while \textit{adaptive} time selection allocates time per input.

We first confirm that the time cost works as expected.
$\mathcal{L}_{\text{time}}$ decreases with increasing $\lamtime$ (Fig.~\ref{fig:time}a), and time used decreases accordingly (Fig.~\ref{fig:time}b).
For both fixed and adaptive schemes, increasing time used increases accuracy;
at matched time used, however, adaptive time selection achieves higher accuracy than the fixed scheme (Fig.~\ref{fig:time}c).

The adaptive model also exhibits sensible behavior on out-of-distribution inputs.
When occlusion is introduced at test time---something the model has never seen during training---time used increases with the proportion of the image occluded (Fig.~\ref{fig:time}d).
% The model spontaneously chooses to compute longer when inputs are degraded.
Examining individual CIFAR-10 images, the model spends the least time on canonical, easy-to-classify examples and the most time on ambiguous or atypical images (Fig.~\ref{fig:time}e).

We compare the model's time allocation with human behavior using CIFAR-10H \cite{peterson_human_2019}, which provides per-image human reaction times and label distributions.
At the category level, model time and human reaction time order the classes similarly (Fig.~\ref{fig:time}f).
Both are fastest on automobiles and slowest on deer, though humans are fast and the model is slow to recognize horses.
At the image level, model time correlates with human reaction time ($\rho = 0.299$, $p < 0.001$; Fig.~\ref{fig:time}g) and with human disagreement, measured as participant response entropy ($\rho = 0.156$, $p < 0.001$; Fig.~\ref{fig:time}h).
These correlations are modest, but the RT correlation survives controlling for image difficulty (partial $\rho = 0.258$; Appendix~\ref{appendix:difficulty-control}) and is not driven by the slowest images (Appendix~\ref{appendix:long-rts}).
A time cost alone, without training on human data, makes the model allocate time in a way that partly tracks human behavior.

\subsection{Breadth vs.\ depth vs.\ time}

We now turn to optimizing all three resource costs jointly.
Accuracy decreases as any of the three costs increases, but the costs interact (Fig.~\ref{fig:breadth-depth-time}a).
Raising the price of time barely lowers accuracy when space is plentiful (0.88 $\rightarrow$ 0.85 from $\lamtime = 0$ to $1$ at $\lambreadth = \lamdepth = 0$) but lowers it much more when space is scarce (0.54 $\rightarrow$ 0.38 at $\lambreadth = \lamdepth = 10^4$).
In other words, time is most valuable to small networks, as expected if it substitutes for space.

To test whether the three resources are interchangeable, we identify, among models with at least 70\% accuracy, the empirical Pareto set---models not dominated by any other model in channels, layers, and time steps used (Fig.~\ref{fig:breadth-depth-time}b).
The 32 Pareto-optimal models span 7.2--39.8 channels, 2--6 layers, and 1.1--4.9 time steps, spreading across every pairwise projection (Fig.~\ref{fig:breadth-depth-time}c).
Breadth, depth, and time are thus fungible: a model can use less of one resource while maintaining accuracy by using more of another.
Stricter thresholds narrow the set, but at $\geq$85\% the 18 Pareto-optimal models still span 21.2--48.0 channels, 3--6 layers, and 1.1--4.9 time steps (Appendix~\ref{appendix:pareto}).

Beyond accuracy, we ask whether models with different resource profiles arrive at different solutions.
We compute error consistency \cite{geirhos_beyond_2020}, which corrects for agreement expected by chance, between all pairs of model configurations (Fig.~\ref{fig:breadth-depth-time}d).
Sorting by $\lamdepth$ reveals a block structure, where shallow models make a consistent set of mistakes that differs from that of deep models, suggesting that depth may change the solution strategy and not only capacity.
Breadth sorting shows weaker structure, and time sorting shows little beyond the diagonal, consistent with additional recurrent steps refining rather than changing the computation.
Similarly, an MDS embedding of pairwise Jensen--Shannon divergences between model outputs (Fig.~\ref{fig:breadth-depth-time}e) is organized by channels, layers, and time used along partially distinct axes beyond accuracy alone.

\subsection{Networks grow with task complexity}
\label{sec:task-complexity}

\begin{figure}[t]
    \centering
    \resizebox{0.98\textwidth}{!}{%
     \begin{tikzpicture}
        % Panel a
        \node[anchor=north west, inner sep=0] (a) at (0, 0)
            {\includegraphics[width=\textwidth]{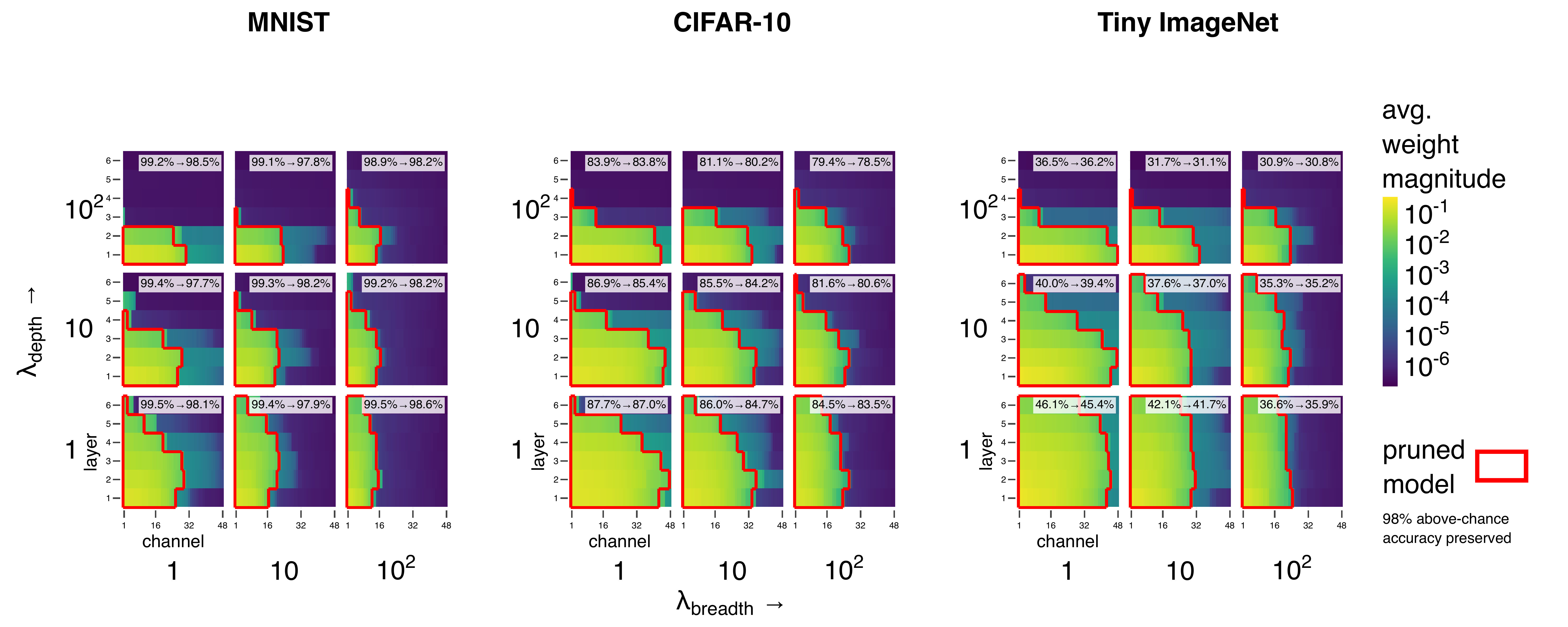}};
            
        % Sample images overlaid on top of the figure
        \node[anchor=south, inner sep=0] (mnist) at ([yshift=-1.18cm, xshift=0.42cm]a.north west -| 0.15\textwidth, 0)
    {\includegraphics[width=0.1\textwidth]{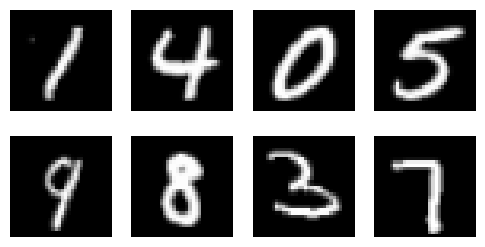}};
    
    \node[anchor=south, inner sep=0] (cifar) at ([yshift=-1.18cm, xshift=-0.48cm]a.north west -| 0.5\textwidth, 0)
    {\includegraphics[width=0.1\textwidth]{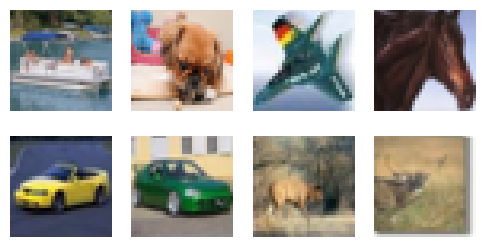}};
    
    \node[anchor=south, inner sep=0] (tiny) at ([yshift=-1.18cm, xshift=-1.28cm]a.north west -| 0.85\textwidth, 0)
    {\includegraphics[width=0.1\textwidth]{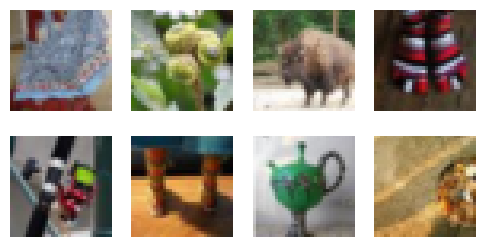}};
        
        % Panel b
        \node[anchor=north west, inner sep=0] (b) at (0, -5.6)
            {\includegraphics[width=0.45\textwidth]{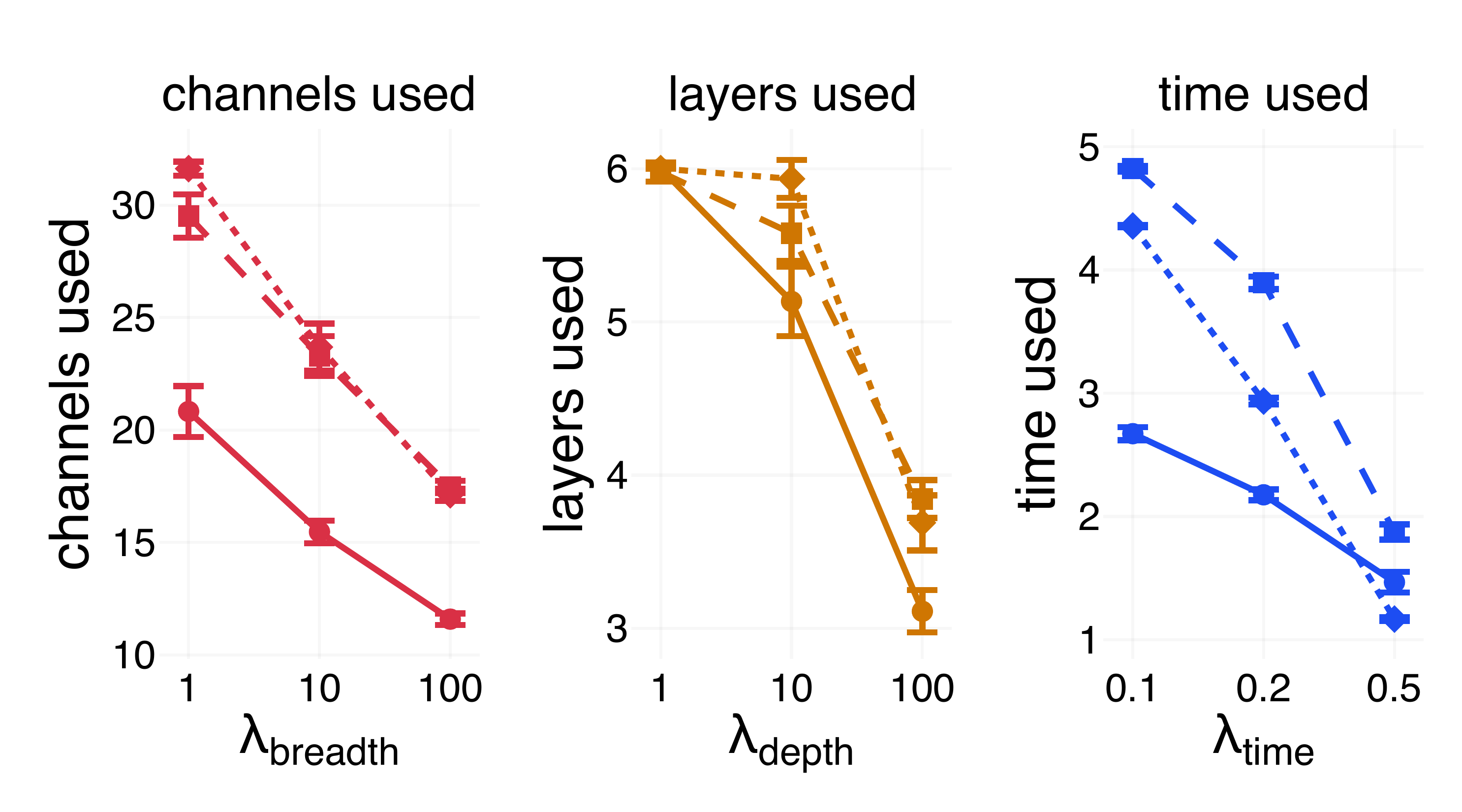}};
        
        % Panel c
        \node[anchor=north west, inner sep=0] (c) at (0.47\textwidth, -5.6)
            {\includegraphics[width=0.55\textwidth]{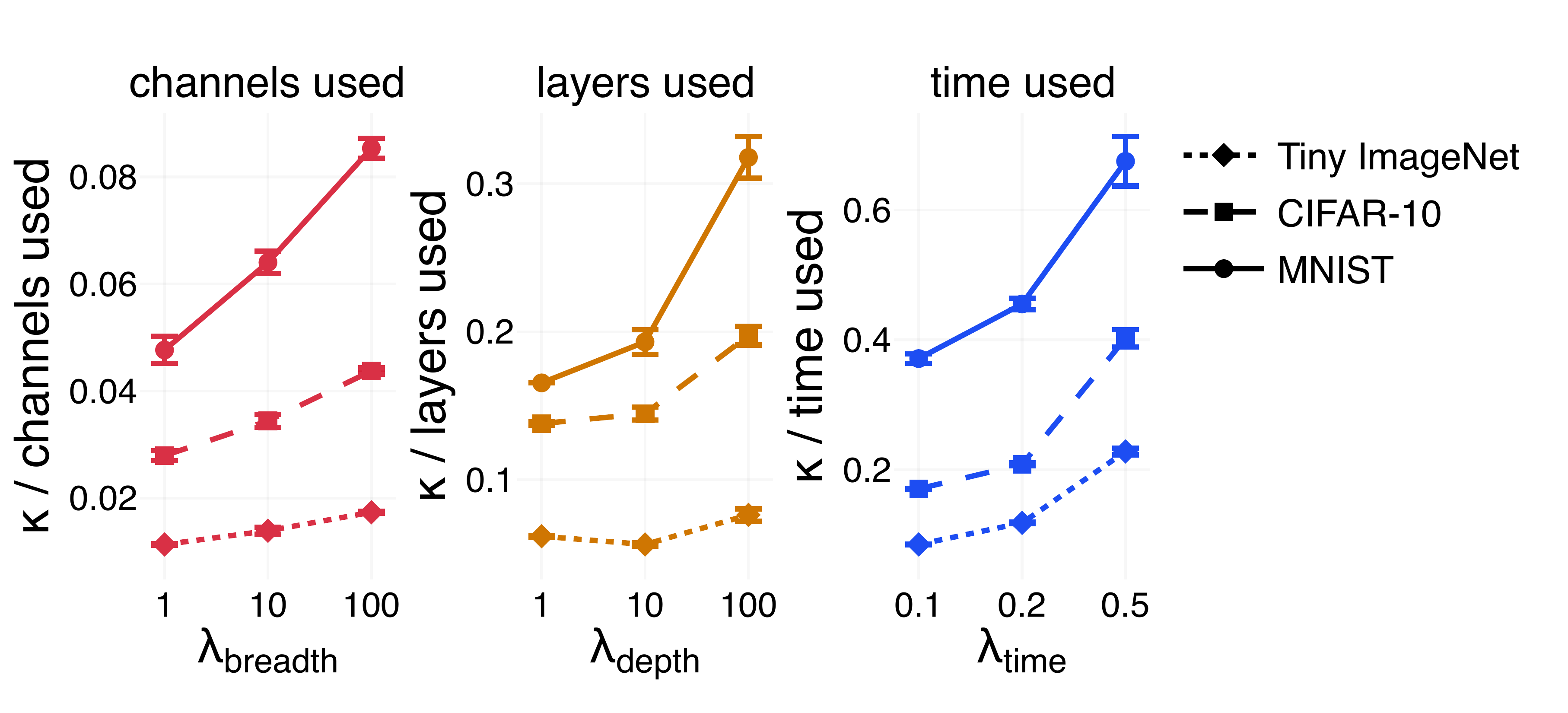}};
    
        % Labels
        \node[font=\normalsize\bfseries] at ([shift={(-0.0cm,-0.3cm)}]a.north west) {a};
        \node[font=\normalsize\bfseries] at ([shift={(-0.0cm,-0.3cm)}]b.north west) {b};
        \node[font=\normalsize\bfseries] at ([shift={(-0.0cm,-0.3cm)}]c.north west) {c};
    \end{tikzpicture}
    }
    \caption{
    \textbf{Task complexity.}
    ~\textbf{a}~Weight magnitude across layers and channels for MNIST, CIFAR-10, and Tiny ImageNet ($\lamtime = 0.1$; one model per panel).
    ~\textbf{b}~Channels, layers, and time used vs.\ resource pressure. CIFAR-10 and Tiny ImageNet networks use more spatial resources than MNIST.
    ~\textbf{c}~Resource efficiency ($\kappa$ / resource used), where $\kappa = (\text{acc} - \text{chance}) / (1 - \text{chance})$.
    MNIST models are most efficient and Tiny ImageNet models least.
    }
    \label{fig:datasets}
\end{figure}

We next test whether networks use more resources as tasks become more complex, training on each dataset under the same grid of resource pressures and comparing at matched $\lambda$ values.
We compare MNIST \cite{lecun1998gradient}, CIFAR-10 \cite{krizhevsky2009learning}, and Tiny ImageNet \cite{le2015tiny} (200 classes), in order of increasing complexity, operationalized as the inverse of our models' average chance-normalized accuracy (Appendix~\ref{appendix:task-complexity}).

The weight magnitude maps (Fig.~\ref{fig:datasets}a) show that natural-image tasks produce larger networks than MNIST.
Under the same resource pressures, MNIST models are sparse while CIFAR-10 and Tiny ImageNet models remain substantially fuller.
Quantitatively, models trained on natural-image tasks use more channels, layers, and time steps than MNIST models for most $\lambda$ settings (Fig.~\ref{fig:datasets}b).
CIFAR-10 and Tiny ImageNet models are similar in size, which partly reflects the normalization of the error cost by $\log K$: at matched $\lambda$, resources are priced against the fraction of the task solved rather than against absolute information (Appendix~\ref{appendix:normalizing-error}).
Measured as $\kappa$ per unit of resource, MNIST models are the most efficient and Tiny ImageNet models the least (Fig.~\ref{fig:datasets}c), although this partly follows from operationalizing complexity through $\kappa$.

\section{Discussion}\label{section:discussion}

We implemented multi-resource cost (MRC) optimization for breadth, depth, and time, showing that the three resources can be traded off against one another and that networks grow with task complexity.
Adaptive time allocation emerges under time pressure and partly tracks human reaction times, suggesting that resource costs can drive aspects of human visual inference.

\paragraph{Limitations.}
Our networks are small-to-intermediate in scale, and whether the same trade-offs hold for larger models and other architectures remains to be tested.
They are also trained with supervised classification on static images, whereas biological vision learns largely from unlabeled, embodied experience.
However, supervised CNNs are strong models of the ventral stream \cite{yamins2014performance, schrimpf2020integrative}, self-supervised models perform comparably to supervised ones \cite{zhuang2021unsupervised, konkle2022self}, and our error term can be replaced by any such objective.
Another limitation is that our resource costs are abstractions that may not map directly onto brains or hardware.
Breadth and depth are priced through rank-based weight magnitudes rather than physical wiring, and time used measures reliance on later steps rather than computation skipped by halting early (Appendix~\ref{appendix:act}).
Furthermore, the model--human comparison is coarse (one model time and one median reaction time per image), and the datasets (MNIST, CIFAR-10, Tiny ImageNet) differ along several axes besides complexity (Appendix~\ref{appendix:task-complexity}).
Finally, we do not benchmark against single-resource methods \cite{wang2018skipnet, liu_learning_2017, graves_adaptive_2016}, as our goal is to study joint trade-offs rather than to maximize performance on any one dimension.

\paragraph{Future directions.}
We plan to extend this approach to energy \cite{ali_predictive_2022, butkus_how_2026} and data, and to resource prices that vary with context, such as a price of time that rises when a predator is near.
Different combinations of resource pressures and task demands can be loosely interpreted as ecological niches, and the solutions that emerge under each may help explain the diversity of neural architectures in nature \cite{laurent_value_2020}, such as the shallower visual hierarchy of mice relative to primates \cite{nayebi2023mouse}.
Development could likewise be modeled as a trajectory through resource price space.
If the resource prices $\boldsymbol{\lambda}$ change as the brain matures, the sequence of architectures that emerges under the changing prices could be compared with the structural and functional changes observed in the developing brain \cite{bethlehem2022brain}.

\paragraph{Conclusion.}
Rather than fixing an architecture or comparing hand-designed ones, we priced breadth, depth, and time jointly and let the architecture emerge through optimization.
This makes it practical to map trade-offs that are costly to explore by hand and to ask which resources can substitute for one another and which solutions are viable under a given budget.
Answering such questions may help explain how brains achieve so much with so little, and how to build AI that does the same.

\clearpage
\section*{Acknowledgments \& Competing Interests}

This work is supported by the funds provided by the National Science Foundation and by DoD OUSD (R\&E) under Cooperative Agreement DBI-2229929 (The NSF AI Institute for Artificial and Natural Intelligence).

We declare no competing interests.

% references
{
\small
\bibliographystyle{plainnat}
\bibliography{references}
}

% appendix
\newpage
\appendix

\section{Code}
\label{appendix:code}

\url{https://github.com/eivinasbutkus/growing-a-neural-network-in-breadth-depth-and-time}

\section{Datasets and Licenses}
\label{appendix:licenses}
We use four publicly available datasets for non-commercial research only and do not redistribute them.
MNIST~\cite{lecun1998gradient} is released under the CC BY-SA 3.0 license.
CIFAR-10~\cite{krizhevsky2009learning} has no explicit license; it is distributed by the University of Toronto (\url{https://www.cs.toronto.edu/~kriz/cifar.html}).
Tiny ImageNet~\cite{le2015tiny} is a subset of ImageNet~\cite{deng2009imagenet} and is subject to the ImageNet terms of access (non-commercial research and educational use).
CIFAR-10H~\cite{peterson_human_2019} (human labels and reaction times for the CIFAR-10 test set) is released under the CC BY-NC-SA 4.0 license (\url{https://github.com/jcpeterson/cifar-10h}).

\section{Compute}\label{appendix:compute}

Each model was trained on a single GPU for approximately 2.5 hours, requiring roughly 3.3 GB of GPU memory at batch size 128. Training was conducted on a university cluster with a mix of NVIDIA GeForce RTX 2080 Ti, A40, and L40 GPUs. The 1,273 models reported in this paper required approximately 3,200 GPU-hours in total. Additional compute was used during model development and preliminary experiments but was not formally tracked.

\section{Experiments}\label{appendix:experiments}

\begin{table}[h]
% \centering
\caption{Summary of experimental configurations. Total models is the product of all variable levels and the number of instances. The breadth vs.\ depth results use the $\lamtime = 0$ slice of the three-way experiment.\\}
\label{tab:experiments}
\small
\begin{tabular}{@{}l l l c r@{}}
\toprule
\textbf{Experiment} & \textbf{Variable} & \textbf{Values} & \textbf{Instances} & \textbf{Models} \\
\midrule
\multirow{3}{*}{Breadth vs.\ depth vs.\ time}
  & $\lambreadth$ & \{0, 1, 10, $10^2$, $10^3$, $10^4$\} & \multirow{3}{*}{3} & \multirow{3}{*}{648} \\
  & $\lamdepth$   & \{0, 1, 10, $10^2$, $10^3$, $10^4$\} & & \\
  & $\lamtime$    & \{0, 0.1, 0.2, 0.3, 0.5, 1.0\}       & & \\
\midrule
\multirow{2}{*}{Time}
  & scheme     & \{fixed, adaptive\}              & \multirow{2}{*}{10} & \multirow{2}{*}{220} \\
  & $\lamtime$ & \{0, 0.1, 0.2, \dots, 1.0\}     & & \\
\midrule
\multirow{4}{*}{Task complexity}
  & dataset       & \{MNIST, CIFAR-10, Tiny ImageNet\} & \multirow{4}{*}{5} & \multirow{4}{*}{405} \\
  & $\lambreadth$ & \{1, 10, $10^2$\}                  & & \\
  & $\lamdepth$   & \{1, 10, $10^2$\}                  & & \\
  & $\lamtime$    & \{0.1, 0.2, 0.5\}                  & & \\
\midrule
\multicolumn{4}{@{}l}{\textbf{Total}} & \textbf{1273} \\
\bottomrule
\end{tabular}
\end{table}

\section{Optimization}

\subsection{Differentiability of the Breadth, Depth, and Time Costs}
\label{appendix:differentiability}

Although the breadth cost involves sorting, it is differentiable with respect to the weights.
The rank-dependent coefficients $(k/C)^\gamma$ are recomputed at every forward pass but treated as constants during backpropagation, so the cost acts as a weighted $\ell_1$ penalty on channel weight magnitudes.
Because the channel ranking changes only when two channels' magnitudes cross, this yields the exact gradient almost everywhere, just as with the absolute value in a standard $\ell_1$ penalty.

The depth and time costs are differentiable by construction.
The depth cost is a weighted $\ell_1$ penalty with fixed layer-dependent coefficients, and the time cost depends on the weights only through the softmax time selection weights.

\subsection{Noise}\label{appendix:noise}

Without internal noise, the magnitude-based breadth and depth costs could be reduced without any loss of function.
The network could shrink the weights of costly units (high-rank channels or deep layers) by an arbitrary factor and compensate downstream, for example with larger readout weights, so that a unit appears unused while still carrying signal.
Inspired by biological neural systems, which contend with multiple sources of noise \cite{faisal_noise_2008}, we add independent Gaussian noise ($\sigma = 0.1$) in two places: to each layer's hidden state at every time step, and to the spatially pooled activations before they enter the side readout and the time-selection MLP.
The second source is needed because spatial pooling averages out the first, which would otherwise leave a high signal-to-noise ratio at the readout.
With noise in place, a signal scaled down toward the noise level is lost, so reducing a unit's weights under resource pressure has a real cost in performance.
This also justifies our magnitude-based pruning (Appendix~\ref{appendix:pruning}).
Channels with small weights contribute little signal relative to the noise, and in practice large portions of the network can be removed without fine-tuning while preserving accuracy (Fig.~\ref{fig:breadth-depth}b), with results robust to the pruning threshold (Appendix~\ref{appendix:pruning-threshold}).

Noise has this effect only if signals cannot be amplified for free before it is added.
Divisive normalization in our model divides by one plus the mean activity, so it compresses strong post-ReLU signals and does not amplify weak ones.
Layers outside the cost terms (the input convolution, the side readout, and the time-selection MLP) are subject to the global weight decay applied to all parameters (Appendix~\ref{appendix:training-details}), which bounds their gain.

The value of $\sigma$ sets the scale against which weight magnitudes are measured, and the network adjusts its weights relative to this floor.
We chose $\sigma = 0.1$ as large enough to close the loophole above yet small enough not to disrupt learning.
Noise is absent for the first 5 epochs and ramped linearly to full strength by epoch 15 (Appendix~\ref{appendix:annealing}); it remains on during evaluation.

\subsection{Annealing Noise and Resource Costs}\label{appendix:annealing}

During an initial warmup period, resource costs are set to zero, allowing the network to first learn useful representations under the error cost alone.
Resource costs are then linearly annealed to their full values over a fixed number of epochs (warmup: 15 epochs for space costs, 20 for time; annealing: 10 epochs for both).
Noise is similarly annealed, starting at zero, then linearly increasing to full magnitude ($\sigma = 0.1$) over epochs 5-15.
This staged schedule prevents the resource costs from interfering with early learning, and ensures the noise floor is established before the costs are fully active.

\subsection{Training Details}\label{appendix:training-details}

All models are trained for 150 epochs using AdamW ($\text{lr}=10^{-3}$, $\beta=(0.9, 0.999)$, weight decay $=0.1$) with mixed-precision training.
The learning rate follows a warmup-hold-cosine schedule: linear warmup for 5 epochs, held constant for 25 epochs (so costs are fully active before decay begins), then cosine decay to $10^{-6}$ over the remaining 120 epochs.
Gradients are clipped to a maximum norm of 1.0.
We chose 150 epochs because we observed no overfitting and only diminishing accuracy gains from 100 to 150 epochs.
Training trajectories also showed no meaningful difference in convergence rate across resource pressures, suggesting that the lower accuracy of strongly constrained models reflects the resource constraints rather than undertraining.

\section{Attribution}
\label{appendix:attribution}
We estimated which spatial regions are important for classification by perturbing the image within a $5 \times 5$ sliding window (stride 1): Gaussian noise ($\sigma=0.05$) was added to the normalized pixel values, which were then clipped to $[0, 1]$. Because inputs are normalized per channel, the clipping dominates, setting darker-than-average pixels to the mean color, so the perturbation acts as a partial occlusion \cite{zeiler2014visualizing}. The score for each patch is the drop in the model's predicted probability for the correct class, $s_{\text{patch}} = p_{c}^{\text{orig}} - p_{c}^{\text{pert}}$, and each pixel's score is the average $s_{\text{patch}}$ over all patches containing it. For Fig.~3f, maps were shifted to be non-negative and normalized to sum to 1 (entropy in nats).

\begin{suppfigure}[H]
% \begin{figure}[h]
    \centering
    \includegraphics[width=0.5\linewidth]{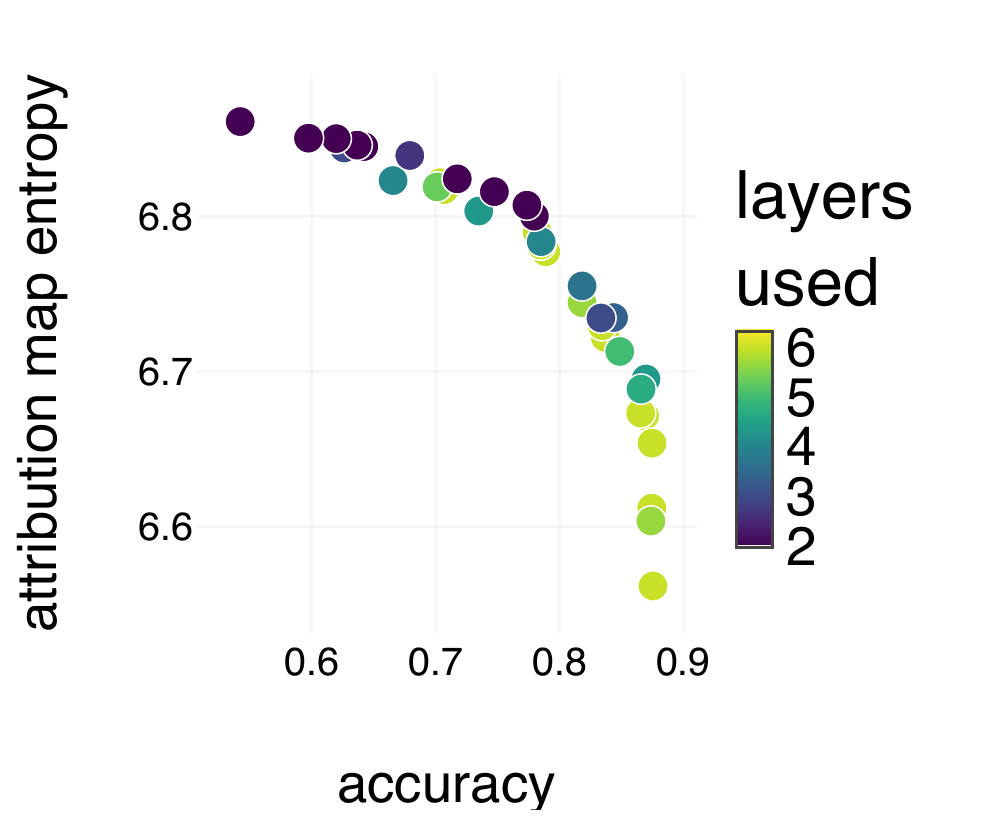}
    \caption{
    Attribution map entropy as a function of accuracy and number of layers used.
    At matched accuracy levels, there is a trend toward higher entropy for shallower models, but the effect is small.
    A more controlled investigation is left for future work.
    }
    \label{fig:entropy-vs-accuracy}
% \end{figure}
\end{suppfigure}

\section{Relation to Adaptive Computation Time (ACT)}
\label{appendix:act}

Our time selection belongs to the same family as adaptive computation time (ACT) \cite{graves_adaptive_2016} and PonderNet \cite{banino2021pondernet}, which define a distribution over recurrent steps and penalize the amount of computation.
A simpler alternative is to train for a fixed number of steps and stop at inference once the output's confidence crosses a threshold \cite{spoerer2020recurrent}, which captures speed--accuracy trade-offs without a time cost during training.

The main difference from ACT is that ACT actually halts.
A halting unit reads out the state at each step, computation stops once the cumulative halting probability reaches $1-\epsilon$, and later steps are never run.
Its ponder cost, $N + R$, is differentiable only through the remainder $R$.
Our model instead runs all $T$ steps and forms its prediction as a softmax-weighted mixture of per-step predictions, with a time cost (Eq.~5) equal to the expected step, which is differentiable with respect to every step's weight.
Time used therefore measures how much the prediction relies on later processing rather than how much computation is skipped.
Converting the learned weights into early stopping at inference is a natural extension.
Because the softmax weights are either free parameters (fixed) or computed from the input (adaptive), the two variants share the same parameterization and time cost, which allows the controlled comparison in Sec.~4.2.

\section{Pruning}

\subsection{Pruning algorithm}
\label{appendix:pruning}

We use an iterative binary search procedure to determine the effective number of layers and channels used by each trained model. The goal is to find the smallest sub-network that preserves 98\% of the model's above-chance accuracy.

We define the accuracy threshold as:
\begin{equation}
    \text{acc}_{\text{threshold}} = \text{acc}_{\text{chance}} + 0.98 \cdot (\text{acc}_{\text{baseline}} - \text{acc}_{\text{chance}})
\end{equation}
where $\text{acc}_{\text{baseline}}$ is the unpruned model's test accuracy and $\text{acc}_{\text{chance}} = 1/K$ for $K$ classes.

Each channel in each layer is assigned a norm equal to its magnitude $\bar{w}$ (Eq.~\ref{eq:breadth}), the same ranking used by the breadth cost during training.
We then perform a binary search in log-space over a norm threshold, and all channels with norm below the threshold are zeroed out.
At each iteration, we evaluate accuracy on the full test set (10,000 images; 4,923 for Tiny ImageNet), using the same noise seed at every iteration so that accuracy differences reflect the pruning mask alone.
If accuracy falls below the threshold, we lower the norm cutoff (preserving more channels).
Otherwise, we raise it (pruning more aggressively).
We run 30 iterations of binary search, which is sufficient for convergence.

The final pruning mask is a binary matrix of shape (layers $\times$ channels). From this mask, we define \emph{layers used} (depth) as the number of layers with at least one surviving channel, and \emph{channels used} (breadth) as the average number of surviving channels per active layer.

\subsection{Resource measures are robust to the pruning threshold}
\label{appendix:pruning-threshold}

We define the resources used by a model through a pruning procedure that finds the smallest sub-network preserving 98\% of the model's above-chance accuracy (Sec.~3.4, Appendix~\ref{appendix:pruning}).
We chose 98\% so that pruned models retain essentially all of their accuracy while leaving some slack for noise in the accuracy estimates.
To test whether our results depend on this choice, we repeated the pruning analysis on the CIFAR-10 breadth--depth models (Sec.~4.1) with the factor $0.98$ in Eq.~6 replaced by $0.99$ and $0.95$ (Table~\ref{tab:pruning-threshold}).
Time used is unaffected by this choice, since it is computed from the time selection weights rather than from pruning.

The resource counts shift only slightly with the threshold: across thresholds, channels used changes by at most 0.64 and layers used by at most 0.34, small compared with the effect of the resource pressures $\lambda$ (from 33 to 7 channels and from 6 to 2 layers).

Small reversals (e.g., slightly more channels at 95\% than at 98\% for
$\lambda_{\text{breadth}} = 0$) arise because internal noise remains on during
evaluation (Appendix~\ref{appendix:noise}), so the accuracy estimates used by the
binary search are themselves noisy.
These differences are within the variability
across instances.

\begin{table}[H]
\centering
\caption{Resources used by CIFAR-10 breadth--depth models ($\lambda_{\text{time}} = 0$) at three pruning thresholds (percentage of above-chance accuracy preserved). For each model instance, values are first averaged over the six values of the other spatial pressure; entries show the mean (standard deviation) across the 3 instances. The 98\% column corresponds to Fig.~3c.}

\label{tab:pruning-threshold}
\small
\begin{tabular}{lccc}
\toprule
 & 99\% & 98\% (default) & 95\% \\
\midrule
\multicolumn{4}{l}{\emph{Channels used}} \\
$\lambda_{\text{breadth}} = 0$ & 33.54 (1.44) & 33.32 (1.35) & 33.96 (0.20) \\
$\lambda_{\text{breadth}} = 1$ & 29.80 (0.26) & 30.19 (0.37) & 29.83 (0.37) \\
$\lambda_{\text{breadth}} = 10$ & 23.40 (0.62) & 23.20 (0.57) & 22.80 (0.23) \\
$\lambda_{\text{breadth}} = 10^2$ & 17.88 (0.24) & 17.68 (0.29) & 17.60 (0.24) \\
$\lambda_{\text{breadth}} = 10^3$ & 12.60 (0.71) & 12.56 (0.71) & 12.61 (0.47) \\
$\lambda_{\text{breadth}} = 10^4$ & 7.00 (0.10) & 7.00 (0.10) & 7.00 (0.10) \\
\midrule
\multicolumn{4}{l}{\emph{Layers used}} \\
$\lambda_{\text{depth}} = 0$ & 6.00 (0.00) & 6.00 (0.00) & 6.00 (0.00) \\
$\lambda_{\text{depth}} = 1$ & 5.94 (0.10) & 5.94 (0.10) & 5.83 (0.00) \\
$\lambda_{\text{depth}} = 10$ & 5.28 (0.10) & 5.17 (0.17) & 4.94 (0.10) \\
$\lambda_{\text{depth}} = 10^2$ & 3.72 (0.25) & 3.72 (0.25) & 3.67 (0.17) \\
$\lambda_{\text{depth}} = 10^3$ & 2.28 (0.10) & 2.28 (0.10) & 2.22 (0.10) \\
$\lambda_{\text{depth}} = 10^4$ & 2.00 (0.00) & 2.00 (0.00) & 2.00 (0.00) \\
\bottomrule
\end{tabular}
\end{table}

\section{Emergent Adaptive Processing}

\subsection{Occlusion experiment}
\label{appendix:occlusion-experiment}

To test whether adaptive models allocate more computation to harder inputs, we occluded CIFAR-10 test images at test time only; models were trained on unoccluded images. Each image was occluded by a single square patch covering a proportion $p \in \{0, 0.05, 0.10, 0.15, 0.20\}$ of the image (side lengths 7, 10, 12 and 14 px on 32×32 inputs). The patch location was sampled uniformly at random over all positions where it fits fully inside the image, independently for each image and model.
Patch pixels were set to zero after per-channel normalization, i.e., to the dataset-mean color (uniform gray).
We evaluated the adaptive-time models trained with $\lamtime$ = 0.2, the intermediate input-dependent time allocation regime (10 seeds), on all 10,000 test images at each occlusion level.
We report mean time used (expected selected timestep, 1–5) per seed; error bars are 95\% t-confidence intervals across the 10 seeds.
As occlusion increased from 0\% to 20\%, accuracy fell from 87.2\% to 67.7\% and time used rose from 2.73 to 2.87 timesteps.

\subsection{Human-model time correlation persists after controlling for image difficulty}
\label{appendix:difficulty-control}

Both humans and the model spend more time on images that are harder to classify (Fig.~4g--h).
The correlation between model time used and human reaction time could therefore reflect image difficulty alone, rather than a finer correspondence in how processing time is allocated across images.
We test this by controlling for a per-image measure of difficulty.

\paragraph{Variables.}
For each of the 10{,}000 CIFAR-10 test images we compute three quantities:
(i) \emph{model time used}, averaged across the 100 adaptive model instances with $\lambda_{\text{time}} > 0$ (10 values of $\lambda_{\text{time}}$ $\times$ 10 seeds; the same models as in Fig.~4e--h);
(ii) human reaction time, the median across the CIFAR-10H judgments of that image
(about 50 per image, excluding attention-check trials)~\cite{peterson_human_2019};
and (iii) \emph{image difficulty}, the entropy of the human response distribution across annotators, computed from the same trials and identical to the measure in Fig.~4h.
Entropy is zero when all annotators choose the same label and increases with disagreement.

\paragraph{Partial correlation.}
We rank-transform all three variables (ties receive average ranks) and apply the standard first-order partial correlation formula to the ranks.
This gives a Spearman partial correlation between model time and human reaction time with difficulty held fixed.
Confidence intervals come from a percentile bootstrap over images (2{,}000 resamples).
The raw correlation is $\rho = 0.299$.
Difficulty is strongly related to human reaction time ($\rho = 0.565$) and more weakly to model time ($\rho = 0.156$; Fig.~4h), so controlling for it removes a large share of the reaction-time variance before the partial correlation is computed.
The partial correlation is $\rho = 0.258$ (95\% CI [0.241, 0.276]), retaining 87\% of the raw correlation (Table~\ref{tab:difficulty-control}).
Using human accuracy (the fraction of annotators choosing the correct label) as the difficulty measure gives nearly the same result ($\rho = 0.260$, 95\% CI [0.243, 0.278]).

\paragraph{Stratified analysis.}
As a check that does not rely on the partial-correlation formula, we split images into strata by entropy and compute the correlation within each.
Because 36\% of images have zero entropy, the quartile cut points are not unique, and the data divide into three non-empty strata (the lowest contains all zero-entropy images together with low-entropy images up to the median).
The correlation is positive in every stratum and strongest among the highest-entropy images.
Finally, restricting the analysis to the 3{,}565 images on which every annotator chose the same label leaves no difficulty variance for the correlation to exploit, yet model time still predicts human reaction time ($\rho = 0.213$).

\begin{table}[h]
\centering
\caption{Spearman correlation between model time used and median human reaction time across CIFAR-10 test images, before and after controlling for image difficulty.}
\label{tab:difficulty-control}
\small
\begin{tabular}{lrl}
\toprule
Analysis & Images & $\rho$ \\
\midrule
Raw correlation & 10{,}000 & 0.299 \\
Partial, controlling for response entropy & 10{,}000 & 0.258 \;[0.241, 0.276] \\
Partial, controlling for human accuracy & 10{,}000 & 0.260 \;[0.243, 0.278] \\
\midrule
Low-entropy stratum & 5{,}511 & 0.212 \\
Intermediate-entropy stratum & 2{,}061 & 0.271 \\
High-entropy stratum & 2{,}428 & 0.305 \\
Unanimous images only (entropy $= 0$) & 3{,}565 & 0.213 \\
\bottomrule
\end{tabular}
\par\smallskip
\footnotesize Brackets: 95\% bootstrap CIs. All correlations $p < 10^{-35}$.
\end{table}

\paragraph{Interpretation.}
Image difficulty, as measured by human disagreement, accounts for only a small part of the correspondence between model and human processing time.
The model's time allocation also tracks image properties that slow humans down even when they agree on the label.
This does not establish that humans and the model use the same mechanism.
Both measures are a single number per image, and human reaction times also include motor and interface components (Appendix~\ref{appendix:long-rts}).

\subsection{Long reaction times do not drive the human-model correlation}
\label{appendix:long-rts}

Median human reaction times in CIFAR-10H are on the order of 1--2\,s, longer than a few additional recurrent cycles in cortex would require.
We attribute much of this to the task interface: participants responded by clicking one of ten labels arranged around the image, so each reaction time includes visual search for the correct label and the mouse movement, which plausibly adds on the order of a second \cite{peterson_human_2019}.
The slowest images may additionally engage multiple fixations, covert shifts of attention, or top-down feedback.
Table~\ref{tab:rt-distribution} shows the distribution of reaction times across all image trials.

\begin{table}[h]
\centering
\caption{Distribution of CIFAR-10H reaction times across all 514,200 image trials
(attention-check trials excluded; 12 trials with negative recorded reaction times omitted).}
\label{tab:rt-distribution}
\small
\begin{tabular}{lrr}
\toprule
Reaction time & Trials & Fraction \\
\midrule
$<$ 1,000 ms      & 35,701  & 6.9\%  \\
1,000--1,500 ms   & 227,832 & 44.3\% \\
1,500--2,000 ms   & 126,721 & 24.6\% \\
2,000--3,000 ms   & 77,273  & 15.0\% \\
$>$ 3,000 ms      & 46,661  & 9.1\%  \\
\bottomrule
\end{tabular}
\end{table}

To test whether the correlation between model time used and human reaction time is carried by the slowest images, we restrict the analysis to images whose median reaction time falls below a threshold, keeping all trials for the retained images so that each per-image estimate remains reliable (Table~\ref{tab:rt-threshold}).
Excluding the 1\% of images with median reaction times above 3\,s leaves the correlation essentially unchanged.
Within the fastest 60\% of images (median below 1.5\,s) the correlation is attenuated, as expected from the restricted range of reaction times, but remains clearly positive ($\rho = 0.150$).
Only 110 images have a median below 1.2\,s, too few for a reliable estimate.

\begin{table}[h]
\centering
\caption{Spearman correlation between model time used and median human reaction time, restricted to images with median reaction time below a threshold.}
\label{tab:rt-threshold}
\small
\begin{tabular}{lrrl}
\toprule
Median RT below & Images & Fraction & $\rho$ \\
\midrule
1{,}200\,ms & 110 & 1.1\% & $-0.044$ ($p = 0.65$) \\
1{,}500\,ms & 5{,}993 & 59.9\% & $0.150$ \\
2{,}000\,ms & 9{,}211 & 92.1\% & $0.249$ \\
2{,}500\,ms & 9{,}767 & 97.7\% & $0.285$ \\
3{,}000\,ms & 9{,}912 & 99.1\% & $0.293$ \\
All images & 10{,}000 & 100\% & $0.299$ \\
\bottomrule
\end{tabular}
\par\smallskip
\footnotesize All correlations except the first: $p < 10^{-30}$.
\end{table}

\section{Pareto Sets at Higher Accuracy Thresholds}

\label{appendix:pareto}
\begin{table}[H]
\centering
\caption{Range of resources used by Pareto-optimal models (Fig.~5b--c) at increasing accuracy thresholds on CIFAR-10.
As the threshold rises, fewer models are Pareto-optimal and the minimum resources used increase, but solutions remain diverse in all three resources.
}
\label{tab:pareto}
\small
\begin{tabular}{lrccc}
\toprule
Accuracy threshold & Pareto models & Channels used & Layers used & Time used \\
\midrule
$\geq$ 70\% & 32 & 7.2--39.8 & 2--6 & 1.10--4.89 \\
$\geq$ 80\% & 21 & 14.3--39.8 & 3--6 & 1.10--4.84 \\
$\geq$ 85\% & 18 & 21.2--48.0 & 3--6 & 1.10--4.89 \\
\bottomrule
\end{tabular}
\end{table}

\section{Task Complexity}

\subsection{Operationalizing task complexity}
\label{appendix:task-complexity}

We operationalize task complexity as the inverse of chance-normalized accuracy, $\kappa = (\text{acc} - \text{chance})/(1 - \text{chance})$, averaged across resource pressures and model instances.
By this measure, complexity increases from MNIST ($\kappa = 0.990$) to CIFAR-10 ($0.796$) to Tiny ImageNet ($0.328$).
The three datasets differ along several confounded axes, including semantic content, number of classes, image statistics, intra-class variability, dataset size, and ambiguity, so this ordering does not isolate which factor drives the increase in resource use; controlled manipulations, such as subsampling Tiny ImageNet to 10 classes, are left for future work.

\subsection{Normalizing the error cost}
\label{appendix:normalizing-error}

Dividing the cross-entropy by $\log K$ (Eq.~\ref{eq:task}) places the error cost on the same scale for every dataset (0 for a perfect model, 1 at chance), while the resource costs depend only on the architecture and are identical across datasets.
A given $\lambda$ therefore prices resources against the same fraction of the task across datasets.
Because $\log K$ is the label entropy for balanced classes, the normalized error can be read as the fraction of label information the model has not yet extracted.
An alternative is to leave the cross-entropy unnormalized, which prices resources per nat of error instead.
Since Tiny ImageNet has more label information to extract ($\log 200 \approx 5.3$ vs.\ $\log 10 \approx 2.3$ nats), its error term would then exert about $2.3$ times more pressure at the same $\lambda$, and its networks would likely use more resources.
Comparisons across datasets at matched $\lambda$ (Sec.~\ref{sec:task-complexity}) are relative to the normalized convention.

\end{document}